\documentclass[twocolumn]{aa}

\usepackage{verbatim}

\bibpunct{(}{)}{;}{a}{}{,} % to follow the A&A style
\usepackage{graphicx}   % Including figure files
\usepackage{txfonts}
\usepackage{array}
\usepackage{hyperref}
\usepackage{enumitem}
\setlist[itemize]{label=-}
\usepackage{newtxtext,newtxmath}
\usepackage{amsmath}    % Advanced maths commands
\usepackage[dvipsnames]{xcolor}
\usepackage{subcaption}

\usepackage{hyperref}  % Add this line

\hypersetup{
    colorlinks=true,
    linkcolor=black,       % Color for internal links
    citecolor=blue, % Change this to your desired citation color
    urlcolor=blue          % Color for URLs
}

\usepackage[splitrule]{footmisc} %% The splitrule option draws a full width rule above the continued part of the footnote as a visual cue to readers.
\usepackage{titlesec} % Adjust spacing before and after section titles

\begin{document} 

    \title{Gravitational waves from cosmic dawn: \\ Tracing cosmic black hole binaries with ET, LGWA, and LISA}

\author{Nazanin Davari\inst{1}\thanks{nazanin.davari@inaf.it; rosa.valiante@inaf.it},
          Rosa Valiante\inst{1,2\ast},
          Alessandro Trinca\inst{3,2,1},
          Raffaella Schneider\inst{5,1,2,6},
          Riccardo Caleno\inst{1,2,5}, 
          Monica Colpi\inst{7,8,12}, 
          Manuel Arca Sedda\inst{8,9,10},
          %Alberto Mangiagli\inst{11},
          Matteo Bonetti\inst{4,8},
          Alessandro Lupi\inst{4,8}, 
          Roberto Decarli\inst{11},
          Alberto Sesana\inst{7,8,12}
          }

   \institute{  % or {\fontsize{6}{4}\selectfont affiliation name}
             {\tiny INAF/Osservatorio Astronomico di Roma, Via Frascati 33, 00040 Monte Porzio Catone, Italy} % 1
        \and
            {\tiny INFN, Sezione Roma1, Dipartimento di Fisica, ``Sapienza'' Universit$\grave{\rm a}$ di Roma, Piazzale Aldo Moro 2, 00185, Roma, Italy} % 2
         \and
            {\tiny Institute for Astronomy, University of Edinburgh, Royal Observatory, Blackford Hill, Edinburgh EH9 3HJ, UK} % 3
        \and
             {\tiny Como Lake Center for Astrophysics, DiSAT, Universit$\grave{\rm a}$ degli Studi dell'Insubria,  via Valleggio 11, 22100, Como, Italy} % 4
        \and
            {\tiny Dipartimento di Fisica, "Sapienza" Universit$\grave{a}$ di Roma, Piazzale Aldo Moro 2, 00185 Roma, Italy} % 5
        \and
            {\tiny Sapienza School for Advanced Studies, Viale Regina Elena 291, 00161 Roma, Italy} % 6
        \and
            {\tiny Dipartimento di Fisica “G. Occhialini”, Universit$\grave{\rm a}$ degli Studi di Milano-Bicocca, Piazza della Scienza 3, I-20126 Milano, Italy} % 7
        \and
            {\tiny INFN, Sezione di Milano-Bicocca, Piazza della Scienza 3, I-20126 Milano, Italy} % 8
        \and    
            {\tiny Gran Sasso Science Institute (GSSI), Via Michele Iacobucci 2, I-67100 L'Aquila, Italy} % 9
        \and    
            {\tiny INFN, Laboratori Nazionali del Gran Sasso, Via Giovanni Acitelli 22, I-67100 Assergi, Italy} % 10
        \and
            {\tiny INAF/Osservatorio di Astrofisica e Scienza dello Spazio
            di Bologna, via Gobetti 93/3, I-40129, Bologna, Italy} % 11
        \and 
            {\tiny INAF,/Osservatorio Astronomico di Brera, via Brera 20, Milano, 20121, Italy} % 12
        } 

   \date{Received date / Accepted date }

% \abstract{}{}{}{}{} 
% 5 {} token are mandatory

%\begin{abstract}
\abstract {
    Next-generation detectors, such as the Laser Interferometer Space Antenna (LISA), the Lunar Gravitational Wave Antenna (LGWA), and the Einstein Telescope (ET), will for the first time probe the high redshift Universe, offering unique insight into the birth, growth, and dynamics of the first black holes (BHs) during their earliest stages of formation.
    We aim to predict merger rates and gravitational wave (GW) signatures of "cosmic" binary BHs 
    forming as a result of galaxy mergers at $z\geq 4$.  
    We investigate how BH seeding, accretion physics, and dynamical delays affect their properties and detectability across cosmic epochs. 
    We used the semi-analytic model Cosmic Archaeology Tool (\texttt{CAT}) to trace the evolution and delayed mergers, driven by dynamical friction, of BH binaries formed from light, medium-weight, and heavy seeds under Eddington-limited (EL) and super-Eddington (SE) accretion prescriptions. We employed the \texttt{GWFish} package to evaluate their GW signals and detectability by LISA, LGWA, and ET.
    Our results show the impact of BH accretion and seeding prescriptions on the properties and distribution of detectable sources. In the EL model, the detected populations are dominated by nearly equal-mass binaries (median mass ratios $\tilde{q} = 0.7$). In contrast, SE growth leads to lower mass ratios for LISA detections ($\tilde{q} \sim 0.2$) and medium ratios for ET and LGWA ($\tilde{q} \sim 0.4$ and $0.3$, respectively). 
   We present the total detection rates predicted under the two accretion scenarios.  The SE model allows BHs to grow faster, transferring a significant fraction of detectable systems from the ET band to the LISA band, compared to the EL model. As a result, the predicted LISA detection rate increases from $\sim 32 \, \rm yr^{-1}$ in the EL case to $\sim 64 \, \rm yr^{-1}$ in the SE scenario, and the ET detection rate reduces from $\sim 64 \, \rm yr^{-1}$ in the EL model to only $\sim 4 \, \rm yr^{-1}$ in the SE scenario. 
   In both scenarios, LGWA yields comparable detection rates ($\sim 21 \, \rm yr^{-1}$ in EL and $\sim 12 \, \rm yr^{-1}$ in SE).  
   The combined information encoded in mass ratios, redshift evolution, and merger rates emerge as a promising diagnostic of early BH growth, and forthcoming multiband GW observations may offer a unique opportunity to disentangle accretion physics and seed formation in the first billion years.
   }  
  
%\end{abstract}


        \titlerunning{Gravitational waves from cosmic dawn} % Use to apply a shorter title in the header
        \authorrunning{N. Davari. et al.} % Use to apply a shorter author list in the header

    \maketitle  % the complete heading of your article

%%%%%%%%%%%%%%%%%%%%%%%%%%%%%%%%%%%%%%%%%%%%%%%%%%
%%%%%%%%%%%%%%%%% BODY OF PAPER %%%%%%%%%%%%%%%%%%
 \section{Introduction}
\label{intro}
Understanding the origin and growth of black holes (BHs) is a key challenge in astrophysics. Observations of luminous quasars at $z \gtrsim 6$ powered by $>10^9 \, \rm M_\odot$ BHs (e.g., \citealt{Fan2023}) and JWST detections of active galactic nuclei (AGNs) at $z > 8$ with inferred BH masses of $10^6{-}10^8\,M_\odot$ \citep[e.g.,][]{barro2023, harikane2023, kocevski2023, larson2023, Leung2023,  labbe2023Nature, Maiolino2024bhs, onoue2023, Ono2023, ubler2023, Yang2023, Bogdan2024, kovacs2024, Maiolino2024z11, Napolitano2025} indicate that massive BHs were already in place within the first billion years of cosmic history. The most distant candidate is now at $z \approx 12.3$ \citep{Chavezortiz2025}.
Moreover, many high-redshift AGNs remain undetected in X-rays, highlighting the difficulty of constraining BH growth through electromagnetic observations alone. Only a few candidates at $z>4$ have tentative X-ray counterparts in Chandra deep field surveys \citep{Goulding2023, Vito2023, Maiolino2024Xray} and a faint hard-band signal emerging only from JWST-AGN stacking \citep{Comastri2025}, challenging traditional methods of AGN identification and growth rate estimation \citep[e.g.,][]{scholtz2024_AGNs, Maiolino2024bhs, larson2023}. 
Future X-ray missions such as newAthena \citep[][]{CruiseNewAthena2025} may detect moderately massive BHs ($M_{\rm BH} \gtrsim 10^6\,M_\odot$), but lower-mass and/or obscured systems will likely remain hidden at $z > 4$ \citep[e.g.,][]{pacucci2015, natarajan2017unveiling, valiante2018observability, valiante2021}.

The existence of early massive BHs implies an earlier population of seed BHs that can form via multiple channels:
(i) remnants of Population III stars, i.e., light seeds \citep[$M_{\rm seed} \sim 10{-}10^3\,M_\odot$; e.g.,][]{MadauRees2001, Abel2002, heger2003, yoshida2008, hirano2014, hirano2015, hosokawa2016, sugimura2020};
(ii) runaway stellar or stellar-mass BH collisions, or tidal disruption events in dense clusters, i.e., medium-weight seeds \citep[$\sim 10^3{-}10^4\,M_\odot$; e.g.,][]{omukai2008,devecchi2012,lupi2014, katz2015, reinoso2018, reinoso2019, boco2020, rizzuto2023tde, Liu2024,Rantala2026}; and (iii) direct collapse of gas in atomic cooling halos, under several environmental conditions, i.e., heavy seeds \citep[$\sim 10^4{-}10^6\,M_\odot$; e.g.,][]{omukai2001, lodato2007, regan2009, hosokawa2012, chon2016, latif2016a, becerra2018, wise2019, lupi2021, mayer2015, haemmerle2019, mayer2019a, chon2025}. 
Explaining the rapid assembly of $>10^8\,M_\odot$ BHs by $z \gtrsim 10$ likely requires massive initial seeds and/or episodes of super-Eddington accretion \citep[e.g.,][]{valiante2016, sassano2021, schneider2023, trinca2024superEdd, Lupi2024a,Lupi2024b,Madau2024, Husko2025, Zana2025}.

Gravitational waves (GWs) provide a complementary and potentially unique probe of BHs formation and growth throughout cosmic time. 
The LIGO–Virgo–KAGRA (LVK) Scientific Collaborations have detected a population of stellar-mass binary black holes (BBHs), forming either in star-forming regions \citep[e.g.,][]{Banerjee2018,Arcasedda2018,Dicarlo2019,Ye2026} or through repeated  mergers of stars or BHs in dense environments \citep[e.g.,][]{Rodriguez2019,Antonini2019,Gerosa2021,Fragione2023,Arcasedda2024}. High-mass events ($>100\,M_\odot$) have also been detected, including GW190521 \citep[][]{Abbott2020} and the most recent BBH merger GW231123 \citep[][]{LVK2025massiveBHB}. 
These high-mass events may have formed via hierarchical mergers or dynamical channels, hinting at complex formation scenarios that may also operate efficiently in the early Universe.  However, current ground-based detectors are limited to BHs with $M \sim 10{-}100\,M_\odot$ at $z \lesssim 2$.
Third-generation ground-based interferometers such as the Einstein Telescope \citep[ET; ][]{Punturo2010, Sathya2012, Maggiore2020, Abac2026} and the Cosmic Explorer \citep[CE; ][]{abbott2017, Reitze2019} will expand the horizon, detecting mergers of $\sim 10$--$10^3\,M_\odot$ stellar- and intermediate-mass BHs, up to $z \gtrsim 10$. 
Space-based observatories, such as the Laser Interferometer Space Antenna \citep[LISA;][]{AmaroSeoane2017, amaroSeoane2023, colpi2024lisa}, scheduled for launch in 2035 and with a mission time of 4 to 8 years, will probe the $10^{-4}$--$10^{-1}\,\mathrm{Hz}$ band, detecting massive BBHs ($\sim 10^4$ --$10^7\,M_\odot$) up to  $z \sim 15-20$.
Proposed observatories such as the Lunar Gravitational Wave Antenna (LGWA; \citealt{Ajith2025}) will target the intermediate decihertz regime, bridging LISA and ET and enabling detection of intermediate-mass BBHs. 
Mapping the demographics of BHs through GWs will provide critical constraints on BH formation, accretion, and early galaxy–BH coevolution. In particular, GWs may be the only way to systematically detect and characterize BH binaries with $M_{\rm BH} < 10^5\,M_\odot$ at high redshift, a regime inaccessible to electromagnetic facilities \citep[e.g.,][]{natarajan2017, valiante2018observability, valiante2021}. Indirect evidence for such intermediate-mass BHs is currently limited to the local Universe via ultra-luminous X-ray sources ($L_{\rm X} \sim 10^{39}{-}10^{41}$ erg/s) in nearby galaxies \citep{Merritt2013} and dynamical studies of star clusters \citep{Noyola2008,vandermarel2010,Abbate2019,Haberle2024}.  

In this work, we investigate the GW signatures of BH binaries formed in the early Universe using the \textsc{Cosmic Archaeology Tool} \citep[\texttt{CAT};][]{trinca2022}. We explore how seed formation scenarios and accretion prescriptions shape the population of merging BH binaries and assess their detectability with ET, LGWA, and LISA. Our goal is to determine whether future multiband GW observations can discriminate among competing models of BH formation in the first billion years of cosmic history.
%The paper is organized as follows. 
We describe the \texttt{CAT} model in Sect.~\ref{sec:CAT}, present the BBH catalogs in Sect.~\ref{sec:cosmic_BBH}, and discuss their GW detectability in Sect.~\ref{sec:detectability}. The model caveats and conclusions are given in Sects.~\ref{sec:discussion} and \ref{sec:conclusions}, respectively.
%%%%%%%%%%%%%%%%%%%%%%%%%%%%%%%%%%%%%%%%%%%%%%%%%%%%%%%%%%%%%%%%%
%%%%%%%%%%%%%%%%%%%%%%%%%%%%%%%%%%%%%%%%%%%%%%%%%%%%%%%%%%%%%%%%%
%%%%%%%%%%%%%%%%%%%%%%%%%%%%%%%%%%%%%%%%%%%%%%%%%%%%%%%%%%%%%%%%%
\titlespacing*{\section}{0pt}{2pt}{2pt}  
\section{The Cosmic Archaeology Tool}
\label{sec:CAT}

In this study, we adopt a semi-analytical approach to leverage its statistical power. \texttt{CAT} is designed to trace the formation and growth of BHs and their host galaxies during the first billion years of cosmic history ($z > 4$).
In the following, we present a brief overview of \texttt{CAT} and refer to \citealt{trinca2022, trinca2024, trinca2024superEdd} for further details. \\
The \texttt{CAT} model runs on top of dark matter halo merger histories, generated analytically with the \textsc{GALFORM} algorithm \citep{Parkinson2008}. We simulated ten merger trees for 11 dark matter halos with masses at $z = 4$ uniformly distributed in $\log(M_{\rm h}/M_\odot) = [9.0, 14.0]$ (bin size 0.5), tracing each backward to $z = 24$. We assumed a minimum halo mass of $\sim 10^6\,M_\odot$ to resolve minihalos ($1200$ K $\leq T_{\rm vir} < 10^4\,\mathrm{K}$), where the first stars and BHs are expected to form. Following \citet{trinca2022}, each merger tree realization was weighted by the Sheth-Tormen halo mass function \citep{sheth2001} so that the resulting BH population represents the ensemble-averaged cosmological halo population rather than the specific realization of the simulated merger trees.
Finally, we adopted a flat Lambda Cold Dark Matter ($\Lambda$CDM) cosmology with $\Omega_\Lambda = 0.685$, $\Omega_m = 0.315$, $h = 0.674$, and $\Omega_b = 0.0483$ \citep{planck2020}.  

% Star formation
Overall, the gas mass budget in each halo is set by depletion processes (star formation, BH accretion, stellar- or AGN-driven outflows) and by replenishment processes (stellar products returned to the diffuse ISM, gas infall from the external medium).
The star formation rate in each halo depends on the available gas mass, $M_{\rm gas}$, and is modeled as $\mathrm{SFR} = f_{\rm cool} \frac{M_{\rm gas} \, \epsilon_{\rm SF}}{\tau_{\rm dyn}}$, where $\tau_{\rm dyn}\equiv (R_{\rm vir}^3 / G M_{\rm h})^{1/2}$ is the dynamical timescale of the halo and $\epsilon_{\rm SF}$, the star formation efficiency, is set to $0.05$, calibrated to match the star formation rate and stellar mass densities at $4\leq z\leq 8$ \citep[see][for details]{trinca2022}. The factor $f_{\rm cool}$ accounts for the reduced cooling efficiency in minihalos ($0 < f_{\rm cool} \leq 1$ according to halo properties, e.g., virial temperature $T_{\rm vir}$, redshift $z$, and ISM metallicity and dust content) and on the intensity of Lyman–Werner (LW; $11.2-13.6$ eV) radiation that dissociates H$_2$ molecules. More massive, atomic-cooling halos are assumed to have $f_{\rm cool}=1$. 
Following each episode of star formation, metal-free or metal-poor environments ($Z < 10^{-3.8}\,Z_\odot$) produce Population~III (Pop~III) stars with $10 \leq m_* \leq 300\,M_\odot$ \citep{debennassuti2014, debennassuti2017}, while enriched halos form Population~II (Pop~II) stars with $0.1{-}100\,M_\odot$ \citep{omukai2001, schneider2002, schneider2012b}.
The Pop~III initial mass function (IMF) was sampled stochastically from a Larson distribution with characteristic mass $m_{\rm ch} = 20\,M_\odot$ \citep{larson1998early}. For Pop~II stars, we adopted a Larson IMF with $m_{\rm ch} = 0.35\,M_\odot$.
%%%%%%%%%%%%%%%%%%%%%%%%%%%%%%%%%%%%%%%%%%%%%%%%%%%%%%%%%%%%%%%%%
%%%%%%%%%%%%%%%%%%%%%%%%%%%%%%%%%%%%%%%%%%%%%%%%%%%%%%%%%%%%%%%%%
%BH formation
\subsection{Seed black holes}
In this work, we follow the formation and evolution of light, medium-weight, and heavy BH seeds within the \texttt{CAT} framework based on local environmental conditions (see Table~\ref{tab:seeding}).
Light seeds form as remnants of Pop~III stars with masses in the ranges $M_{\rm seed} = [40{-}140]$ and $[260{-}300]\,M_\odot$ \citep[][]{heger2002}. These arise in metal-free or metal-poor minihalos and atomic cooling halos (ACHs; $T_{\rm vir} > 10^4\,\mathrm{K}$) exposed to weak LW radiation fields, $J_{\rm LW} < J_{\rm crit}$, with $J_{\rm crit} = 300 \times 10^{-21}\,\mathrm{erg\,s^{-1}\,cm^{-2}\,Hz^{-1}\,sr^{-1}}$. Among the remnants formed in a given Pop~III starburst, the most massive BH is selected as the seed and assumed to migrate to the galactic nucleus \citep[see][]{trinca2022, trinca2024}.

Heavy seeds, with $M_{\rm seed} = 10^5\,M_\odot$, form via direct collapse in metal-poor ACHs \citep[$Z < 10^{-3.8}\,Z_\odot$; ][]{ChonOmukai2020}, provided they host a large gas reservoir ($M_{\rm gas} > 10^6\,M_\odot$) and are exposed to strong LW radiation fields ($J_{\rm LW} > J_{\rm crit}$). These halos must also be only slightly affected by photoheating feedback.

In this work, medium-weight seeds are introduced as a new feature of the model. We assigned $M_{\rm seed} = 10^3\,M_\odot$ to halos with intermediate metallicity, $10^{-3.8} < Z/Z_\odot < 10^{-2.5}$, and  similarly exposed to strong LW fluxes \citep[see][]{sassano2021}.
The complete set of seeding conditions, including the thresholds in $T_{\rm vir}$, $M_{\rm gas}$, $Z$, and $J_{\rm LW}$, is summarized in Table~\ref{tab:seeding}. Further details and a critical discussion of the physical assumptions can be found in \citet[][]{valiante2016, sassano2021, trinca2022}.
After formation, we assume that the seed BHs grow through gas accretion and BH–BH mergers.
%%%%%%%%%%%%%%%%%%%%%%%%%%%%%%%%%%%%%%%%%%%%%%%%%%%%%%%%%%%%%%%
\begin{table*}[!h]
    \caption{Seeding criteria adopted for the three BH seed formation channels.}
    \label{tab:seeding}
    \centering
    \begin{tabular}{|c|c|c|c|c|c|}
    \hline
         & & & & &\\
        { seed channel} & { $M_{\rm seed}/M_\odot$} & {$T_{\rm vir}/ K$} & { $M_{\rm gas}/M_\odot$} & { $Z/Z_\odot$} &  { $J_{\rm LW}/J_{21}$}\\
        %& & & & &\\
        \hline
         light & $[40, 140]$; $[260, 300]$ & $>1200$ & $>300$ & $<10^{-3.8}$ & $< 300$\\
         medium-weight & $10^3$ & $>10^4$ & $>10^6$ & $[10^{-3.8}, 10^{-2.5}]$ & $>300$\\
         heavy & $10^5$ & $> 10^4$ & $>10^6$ & $<10^{-3.8}$ & $>300$\\
         \hline
        \end{tabular}
    \tablefoot{The three seed formation channels are light,  medium-weight, and heavy (first column). Columns list the seed mass ($M_{\rm seed}$), minimum virial temperature ($T_{\rm vir}$), required gas reservoir ($M_{\rm gas}$), ISM metallicity ($Z$), and LW flux ($J_{\rm LW}$) in units of $J_{21} = 10^{-21}\,\rm erg\,s^{-1}\,cm^{-2}\,Hz^{-1}\,sr^{-1}$ (see \citealt{sassano2021} for a thorough discussion on these seeding conditions).    }        
\end{table*}
%%%%%%%%%%%%%%%%%%%%%%%%%%%%%%%%%%%%%%%%%%%%%%%%%%%%%%%%%%%%%%%%%%
\subsection{Black hole accretion}

We adopted two accretion models following \citet{trinca2022}.
%\paragraph{The Eddington-limited growth.}
In the Eddington-limited (EL) scenario, the accretion rate is computed using the Bondi formula for spherical accretion \citep[][]{bondi1952}: 
\begin{equation}
    \dot{M}_{\rm accr}=
    %\dot{M}_{\rm BHL} = 
    \alpha \, \frac{4 \pi G^2 M_{\rm BH}^2 \rho_{\rm gas}(r_A)}{c_s^3},
    \label{eq:BHL}
\end{equation}
where $c_s$ is the sound speed, $\rho_{\rm gas}(r_A)$ is the gas density within the Bondi radius $r_A = 2 G M_{\rm BH} / c_s^2$, { computed assuming a singular isothermal sphere with a flat core}, and $\alpha$ accounts for the unresolved gas density near the BH \citep[][]{tanaka2009, dimatteo2012, schaye2015}. We adopt $\alpha = 90$ to match the properties of $z \sim 6{-}7$ quasars \citep{trinca2022}. 
Accretion is capped at the Eddington rate $\dot{M}_{\rm Edd} = L_{\rm Edd} / \epsilon_{\rm r} c^2$ {where $\epsilon_{\rm r}$ is the radiative efficiency set to 0.1}.

In the Super-Eddington (SE) scenario, in addition to Bondi accretion (for which we assumed $\alpha =1$), major galaxy mergers  defined in terms of the stellar mass ratio \citep{hopkins2010mergers, newman2012can} $\mu_\ast = M_{\ast,2}/M_{\ast,1} > 0.1$, trigger short bursts of accretion exceeding the Eddington limit:
\begin{equation}
    \dot{M}_{\rm accr} = \frac{\epsilon_{\rm BH} \, M_{\rm gas}}{\tau_{\rm accr}},
    \label{eq:SEaccr}
\end{equation}
where $M_{\rm gas}$ is the gas mass of the remnant galaxy. We fixed $\epsilon_{\rm BH} = 0.017$ and $\tau_{\rm accr} = 10$ Myr (see \citealt{trinca2022} for more details). Star formation is similarly enhanced on the same timescale:
$
\mathrm{SFR} = \frac{\epsilon_{\rm SF} \, M_{\rm gas}}{\tau_{\rm accr}},
$
using the same $\epsilon_{\rm SF}$ as in the EL model, with $f_{\rm cool}=1$.

In both models, the BH mass grows as
$
{\dot M}_{\rm BH} = (1 - \epsilon_{\rm r})\, \dot{M}_{\rm accr},
$
where the radiative efficiency is fixed at $\epsilon_{\rm r} = 0.1$ in the EL model and calculated from \citet{madau2014a} in the SE case, assuming a BH spin $a = 0.572$.
This spin parameter is only used to compute the radiative efficiency entering the BH growth model, ensuring that $\epsilon_{\rm r}$ approaches $0.1$ when the accretion rate drops below the Eddington limit, consistent with the value adopted in the EL model.

In the EL scenario, light seeds grow inefficiently and the high-mass end of the BH population is dominated by heavy seeds. In contrast, the SE model allows all seeds to grow rapidly, producing a more mixed and massive population \citep[see][for a thorough discussion]{trinca2022}.
%%%%%%%%%%%%%%%%%%%%%%%%%%%%%%%%%%%%%%%%%%%%%%%%%%%%%%%%%%%%%%%%%%

\subsection{Black hole mergers}
\label{sec:dynamics}

In previous work employing \texttt{CAT}, two BHs were assumed to instantaneously form a binary system and coalesce in major halo-halo merger events \citep[][]{trinca2022}. 
In this work, we implement a new prescription, including a delay in the BH-BH merger event, defined as the time-scale required to form a bound binary in the aftermath of galaxy collisions.

Following a major galaxy merger, the most massive BH (primary, $M_1$) is placed at the center of the remnant galaxy and allowed to accrete. The secondary ($M_2$) is assumed to orbit in the outskirts, unable to accrete gas.
Following \citealt{volonteri2020} we computed the timescale due to stellar dynamical friction as:

\begin{equation}
    \label{eq:dynFricTime}
    t_{\rm df}= 19 \, {\rm Gyr} \bigg( \frac{r}{5 \, \rm kpc}\bigg)^{2} \bigg( \frac{\sigma_{\ast}}{200 \, \rm km s^{-1}} \bigg) \bigg( \frac{M_{2}}{10^{8} {\rm M_{\odot}}} \bigg)^{-1} \frac{1}{\Lambda},
\end{equation}
\noindent
where $\Lambda = {\rm ln}(1+M_{\rm star}/M_{2})$ is the Coulomb logarithm, $M_{\rm star}$ is the total (post-merger) stellar mass, and $\sigma_\ast = (0.25 G M_{\rm star}/R_{\rm eff})^{1/2}$ is the stellar velocity dispersion.
We set the initial separation of the two BHs at $r=R_{\rm eff}$, where the effective radius, $R_{\rm eff}$, defined as the half-mass radius of the merger remnant galaxy, is computed using the observed mass-size relation for late-type galaxies from \cite{shen2003size} and accounting for the redshift evolution of the galaxy radius following \cite{vanDerWell2014}:

\begin{equation}
\label{eq:effectiveRadius}
    R_{\rm eff}= 0.1 \, {\rm kpc} \, \bigg( \frac{M_{\rm star}}{\rm M_\odot}\bigg)^{0.14} \bigg( 1+ \frac{M_{\rm star}}{3.98\times 10^{10} \, \rm M_\odot} \bigg)^{0.25} (1+z)^{-0.75}.
\end{equation}

This prescription provides a physically motivated estimate of the characteristic stellar scale over which dynamical friction governs the initial pairing of the secondary BH. Similar galaxy-scale prescriptions are widely adopted in semi-analytical models of MBH dynamics, although modeling the kpc-scale BH pairing phase remains one of the largest sources of uncertainty in predicting massive BH merger rates \citep[e.g.,][]{barausse2020, Izquierdo2022, izquierdoVillalba2026, Singh2026}. For example, at large initial separations, the secondary BH is expected to remain embedded within its host galaxy nucleus, increasing the effective mass participating in dynamical friction and consequently shortening the sinking time \citep[e.g.,][]{Singh2026}.

We assumed that BHs merge at the end of the pairing phase, neglecting additional delays from hardening or GW emission. This is justified as dynamical friction is typically the dominant timescale in low-mass systems \citep[see e.g.,][and sec.~\ref{sec:discussion}]{volonteri2020}. If a new BH enters the system before coalescence, it replaces the existing secondary if its $t_{\rm df}$ is shorter. In mergers involving multiple BHs, we retain only the pair with the shortest $t_{\rm df}$ { often corresponding to the heavier component masses}. 
This is supported by three-body simulations showing that the least massive BH is typically ejected \citep[][]{bonetti2016, bonetti2018a, bonetti2018b}, consistent with the short interaction timescales found by \citet{Fastidio2025}.
%%%%%%%%%%%%%%%%%%%%%%%%%%%%%%%%%%%%%%%%%%%%%%%%%%%%%%%%%%%%%%%%%%
% Stellar and AGN feedback
\subsection{Feedback from stars and accreting black holes}
\label{Feedback from stars and accreting BHs}

In the \texttt{CAT} framework, stellar and AGN feedback is modeled as energy-driven outflows on galactic scales, triggered by supernova (SN) explosions and BH accretion. This mechanical feedback regulates the gas budget within each galaxy and contributes to the enrichment of the external medium \citep[][]{trinca2022}.
The total gas outflow rate is the sum of SN- and AGN-driven components: 
$\dot{M}_{\rm ej}=\dot{M}_{\rm ej,SN}+\dot{M}_{\rm ej,AGN}$, 
where 
\begin{equation}
{\small
\begin{array}{l}
%\dot{M}_{\rm ej} = \dot{M}_{\rm ej,SN} + \dot{M}_{\rm ej,AGN}, \\[8pt]
\dot{M}_{\rm ej,SN} = \dfrac{2 E_{\rm SN} \, \epsilon_{\rm w,SN} \, R_{\rm SN}(t)}{v_{\rm e}^2}, \quad 
\dot{M}_{\rm ej,AGN} = \dfrac{2 \, \epsilon_{\rm w,AGN} \, \epsilon_{\rm r} \, \dot{M}_{\rm accr} \, c^2}{v_{\rm e}^2}.
\end{array}
}
\label{eq:gas_outflow}
\end{equation}
\begin{figure}
    \centering
    \includegraphics[width=1.02\linewidth]{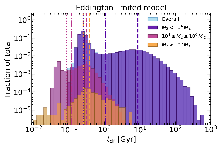}
    \includegraphics[width=1.02\linewidth]{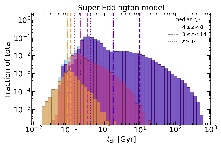}
    \caption{Distribution of BH pairs as a function of their dynamical friction timescales in the EL (top) and SE (bottom) models in different bins for the secondary BH mass. Vertical lines indicate the mean and median values for different mass ($M_2$) and redshift bins.
    }
    \label{fig:tdf}
\end{figure}
Here, $v_{\rm e}=\sqrt{2 G M/R_{\rm vir}}$ is the halo escape velocity, while $E_{\rm SN}$ and $R_{\rm SN} (t)$ are the explosion energy per SN and the total SN explosion rate.
The SN wind efficiency is set to $\epsilon_{\rm {w, SN}}=1.6\times 10^{-3}$ in both the EL and SE models, following \citet{trinca2022}. In this work, the wind efficiency of the AGN is $\epsilon_{\rm w,AGN}=10^{-3}$, a factor of 2.5 lower than previous implementations \citep{trinca2022}, to preserve agreement with the observational constraints used in the original \texttt{CAT} calibration \citep[][see Appendix \ref{appB:recalibration}]{trinca2022}.
Indeed, the introduction of finite dynamical-friction delays modifies the timing of SMBH assembly and, therefore, requires a recalibration of the small set of free parameters controlling BH growth and AGN feedback. 
The resulting value, $\epsilon_{\rm w,AGN}=10^{-3}$, remains fully consistent with the range commonly adopted in current semi-analytical and cosmological models of BH-galaxy  coevolution \citep[e.g.,][]{dave2019, Izquierdo2022}.

%%%%%%%%%%%%%%%%%%%%%%%%%%%%%%%%%%%%%%%%%%%%%%%%%%%%%%%%%%%%%%%%%%
%%%%%%%%%%%%%%%%%%%%%%%%%%%%%%%%%%%%%%%%%%%%%%%%%%%%%%%%%%%%%%%%%
%%%%%%%%%%%%%%%%%%%%%%%%%%%%%%%%%%%%%%%%%%%%%%%%%%%%%%%%%%%%%%%%%
\section{Cosmic black hole binaries and their mergers in the GW universe}
\label{sec:cosmic_BBH}

As discussed above, the growth of the first BHs is governed by hierarchical mergers and gas accretion within evolving halos. We define "cosmic" BBHs as binaries forming during the hierarchical assembly of galaxies. 

Figure~\ref{fig:tdf} shows the distribution of dynamical friction timescales $t_{\rm df}$ for BH pairs, computed according to Eq.~\ref{eq:dynFricTime}.
Histograms are divided into secondary BH mass bins, as labeled in the figure. 
The median dynamical friction timescale ranges from $\sim 0.3$ ($0.5$) Gyr if BH pairs form at $z\geq14$ to $\sim 12$ ($20$) Gyr for BHs pairing by $4\leq z<8$ in the EL (SE) scenario.
As expected, more massive BHs sink faster toward the center, regardless of the accretion model.
Low-mass BH pairs ($M_2<10^4 \, \rm M_\odot$) forming at $z<8$ have median $t_{\rm df}\sim 8-10$ Gyr %in both scenarios 
whereas those forming at earlier epochs evolve more rapidly, merging by $z=4$ ($t_{\rm df}<2$ Gyr). More massive systems, by contrast, have median $t_{\rm df}<0.5$ Gyr across all redshifts.
As a consequence of the dynamical delay and multiple BH encounters (see Sect.~\ref{sec:dynamics}), $\sim 20\%$ of BH pairs form bound binaries and merge by $z=4$, with merger times spanning $10^{-2} < t_{\rm df}/{\rm Gyr} < 1.5$.
The fraction increases to $\sim 40\%$ if systems with $t_{\rm df}$ shorter than Hubble time are included, which would merge at $0<z<4$, assuming that the dual BH system remains unperturbed during the dynamical friction phase.

\begin{figure*}%[htbp]
\centering
\includegraphics[width=0.9\textwidth]{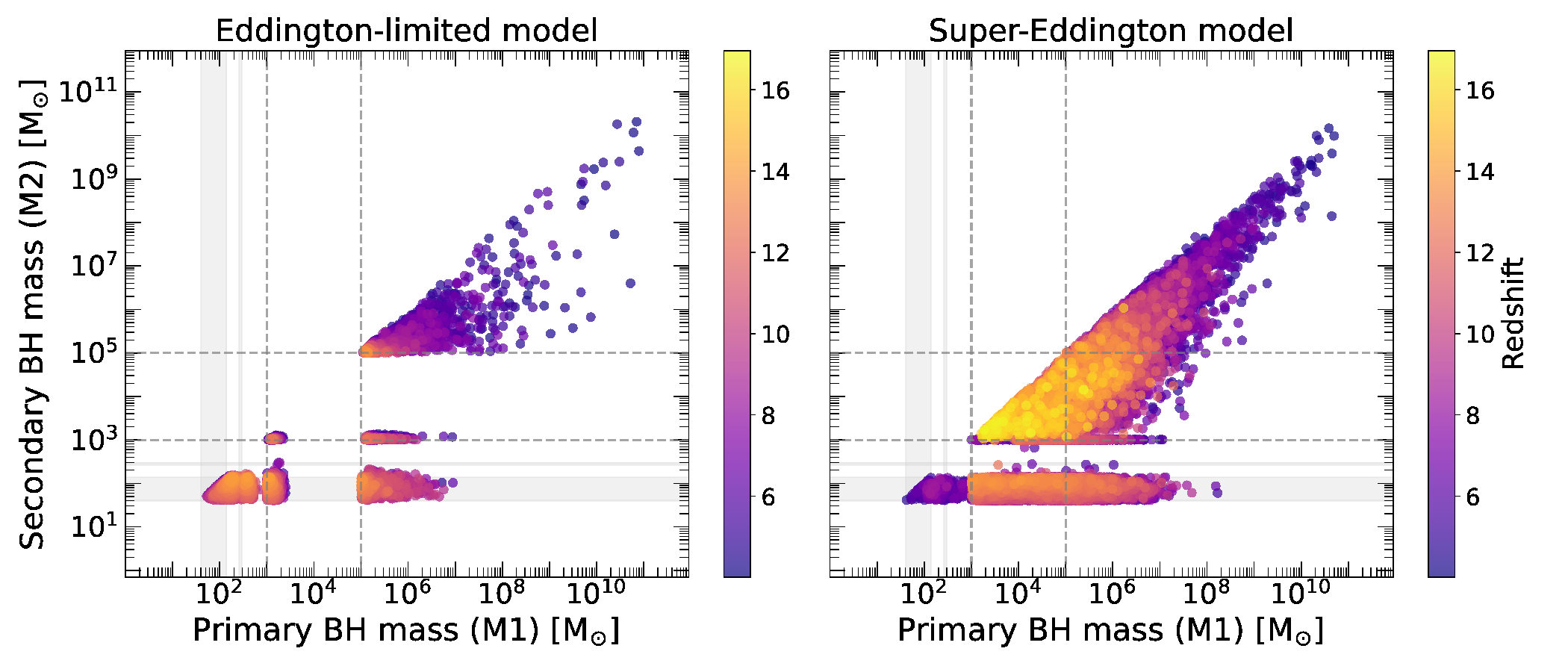} 
\caption{Primary (most massive) vs. secondary (least massive) BH masses in merging binaries, color-coded by redshift, for the EL (left) and SE (right) models. Gray shaded areas mark the light seed mass ranges ($[40, 140]$ and $[260, 300]\,\mathrm{M_\odot}$); dashed lines indicate the adopted masses for medium-weight ($10^3\,\mathrm{M_\odot}$) and heavy ($10^5\,\mathrm{M_\odot}$) seeds. 
}
\label{fig:gwtc_el}
\end{figure*}

The masses of the primary ($M_1$) and secondary ($M_2$) BHs in our cosmic BBHs in the EL and SE scenarios are shown in Fig.~\ref{fig:gwtc_el}. The visible gaps in the EL model are a direct consequence of the adopted BH seeding prescription. %, rather than an effect of the merger-tree sampling. 
In the current implementation of \texttt{CAT}, medium-weight and heavy BH seeds are assigned fixed birth masses ($10^3$ and $10^5 \, \rm M_\odot$, respectively), while light seeds populate two narrow mass intervals according to the adopted Pop III remnant model. Because BH growth is relatively inefficient in the EL scenario, many binaries preserve the imprint of these discrete seed masses, producing visible structures in the BBH mass distribution. 
This behavior is a natural extension of the effect already discussed by \citet{trinca2022}, where the accumulation of BH around the adopted heavy-seed mass in the BH mass function was shown to originate from the discrete seeding prescription. In the present work, the introduction of medium-weight seeds adds a second characteristic seed mass, while the inefficient growth of light BH seeds in the EL scenario preserves the imprint of these discrete birth masses in the BBH population.
This effect is further enhanced by our assumption that secondary BHs do not accrete gas. Without accretion, secondary BHs remain close to their initial seed masses, whereas gas accretion would promote their growth toward intermediate masses and partially smooth these structures. The impact of this assumption on the BBH population is further discussed through the resulting mass-ratio (q) distribution in the Appendix~\ref{appA:mass_ratio}.

%%%%%%%%%%%%%%%%%%%%%%%%%%%%%%%%%%%%%%%%%%%%%%%%%
%%%%%%%%%%%%%%%%%%%%%%%%%%%%%%%%%%%%%%%%%%%%%%%%%

In this work, we assess the detectability of cosmic BBHs mergers predicted by \texttt{CAT} using the \texttt{GWFish} software package \citep{dupletsa2023gwfish}. 
We consider a set of next-generation GW observatories probing complementary frequency bands: LISA in the millihertz, LGWA in the decihertz, and ET above a few hertz. This combination provides illustrative coverage of the GW spectrum from $10^{-4}$ to $\sim 10^2$ Hz. 

We first evaluated the detectability of sources in each band separately and then assess the potential for multiband sources, namely mergers that could, in principle, be detected by more than one detector, highlighting the prospects for coordinated, sequential and/or archival observations across different frequency bands. In this context, the term multiband refers to a conceptual synergistic use of detectors operating across complementary frequency ranges, aimed at probing long-lived signals (see Fig.~\ref{fig:tdf}), assuming a representative observation duration of 4 years.

For each BBH, the GW signal is modeled using the inspiral-merger-ringdown (IMR) \texttt{IMRPhenomXHM} waveform model \citep{garcia2020}, which includes the full inspiral-merger-ringdown evolution of quasi-circular, non-precessing BBHs with aligned spins. 
The model incorporates higher-order modes beyond the dominant quadrupole $(\ell, m)=(2,2)$, including $(2, 1)$, $(3, 3)$, $(3, 2)$, $(4, 4)$. 
The \texttt{IMRPhenomXHM} model is calibrated on hybrid waveforms valid down to $q=0.001$. We therefore restrict our analysis to the \texttt{CAT} merging binaries with $q > 0.001$, which we define as the intrinsic population of \texttt{CAT}, thus excluding extreme mass ratio inspirals (EMRIs). The latter represent $<13\%$ (EL) and $<20\%$ (SE) of the total population. In this selection, the \texttt{CAT} catalogs count $N_{merge}=57958$ systems in the EL model and $N_{merge}=61484$ in the SE case.\\
The signal-to-noise (S/N) ratio of our synthetic BBHs is computed by integrating the GW signal accumulated over a nominal observing time of 4 yr for each detector. 
Throughout this work,, we consider an event to be detected when its $\mathrm{S/N} \geq 8$. 

We consider the set of parameters summarized in Table~\ref{tab:priors}, which includes the component masses ($m_1$, $m_2$), the luminosity distance ($d_L$), the inclination angle ($\theta_{\rm jn}$), the polarization angle ($\psi$), the coalescence phase ($\phi_c$), the sky location $(ra, dec)$, the time of merger in geocentric coordinates $t_c$, and spins ($a_1$, $a_2$). The mass and redshift are taken from the \texttt{CAT} output catalogs, while the others are randomly assigned. We assume non-spinning BHs, but this should have only a minor impact on the results\footnote{While adopting moderate BH spins ($a_1=a_2=0.5$) slightly affects the inferred detection rates and S/N distributions, the key findings and conclusions drawn from the non-spinning analysis remain consistent.}. 

This assumption refers exclusively to the generation of the GW waveform and does not affect the intrinsic BBH population predicted by \texttt{CAT}. Its impact on detectability is expected to be significantly smaller than on parameter estimation \citep[e.g.,][]{klein2016, Babak2021}.

We used ET-D \citep{Branchesi2023}, LGWA-Si \citep{Ajith2025}, and LISA \citep{Babak2021} power spectral densities (PSDs) available in the \texttt{GWFish} repository. 

%%%%%%%%%%%%%%%%%%%%%%%%%%%%%%%%%%%%%%%%%%%%%%%%%%%%%%%%%
\begin{table}[!h]
\caption{Sampling parameters for the BBH populations.}
\label{tab:priors}
\centering
{\small
    \begin{tabular}{|c|c|}
    \hline
    \textbf{Parameter} & \textbf{Source/Sampling} \\
    \hline
    Redshift & from \texttt{CAT}  \\
    %\hline
    $m_1, m_2$ ($M_\odot$) & from \texttt{CAT} ($m_1 \geq m_2$) \\
    %\hline
    $q = m_2/m_1$ & $q \geq 0.001$ \\
    %\hline
    $d_L$ & convert $z$ via \textsc{astropy.Planck18} \\
    %\hline
    $\cos\theta_{\rm jn}$ (rad) & uniform in [-1,1] \\
    %\hline
    ra: $\alpha$ (rad) & uniform in [0,2$\pi$] \\
    %\hline
    dec: $\sin\delta$ (rad) & uniform in [-1,1]\\
    %\hline
    $\psi$ (rad) & uniform in [0,$\pi$] \\
    %\hline
    $\phi_c$ (rad) & uniform in [0,2$\pi$] \\
    %\hline
    $t_c$ (sec) & 4-year period \\
    %\hline
    $a_1, a_2$ & non-spinning \\
    \hline
    \end{tabular}
    }
\end{table}
%%%%%%%%%%%%%%%%%%%%%%%%%%%%%%%%%%%%%%%%%%%%%%%%%%%%%%%%%
For each merging BBH, in both the EL and SE models, we performed ten independent realizations, each based on a randomly selected set of parameters drawn from the distribution described above.
Unless otherwise specified, all quoted quantities and figures refer to the median over these runs.
For every simulated BBH merger, we estimated the S/N, applying a conservative detection limit of $\mathrm{S/N} \geq 8$ per instrument. Approximately $32\%$ of the intrinsic \texttt{CAT} population satisfies this criterion in both accretion scenarios, corresponding to a median sample of 18464 (19697) detectable GW sources ($q > 0.001$ and $\mathrm{S/N} \geq 8$) drawn from the EL (SE) \texttt{CAT} catalog.
%%%%%%%%%%%%%%%%%%%%%%%%%%%%%%%%%%%%%%%%%%%%%%%%%
%%%%%%%%%%%%%%%%%%%%%%%%%%%%%%%%%%%%%%%%%%%%%%%%%%%%%%%%%%

\section{Detectability of gravitational wave sources}
\label{sec:detectability}
\begin{figure*}[htbp]
\centering
\includegraphics[width=\textwidth]{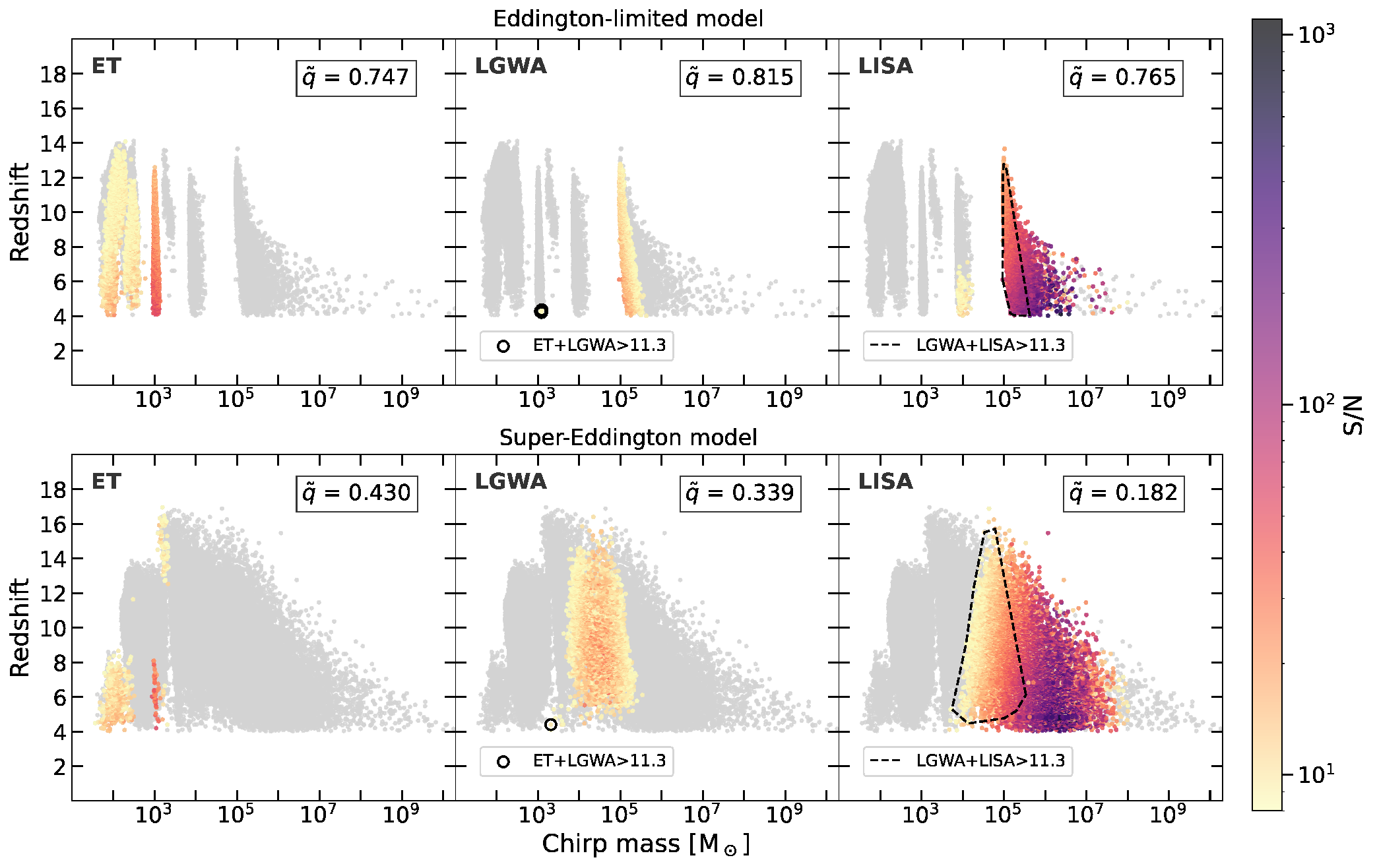} 
\caption{Merger redshift as a function of the chirp mass in the EL (top) and SE (bottom) accretion scenarios. Gray points represent the full population of merging BBHs, while colored points denote detectable sources, shaded by S/N, over 4 yr of observations with ET (left), LGWA (middle), and LISA (right). The median mass ratio ($\tilde{q}$) of the detectable systems is shown in each panel. In the EL model, detections are dominated by nearly equal-mass binaries ($\tilde{q} \approx 0.7-0.8$), whereas the SE scenario favors more asymmetric mergers. The mass and redshift range that would be accessible by 3G detectors depend on the accretion mode. 
The solid circles and dashed lines enclose potential ET$+$LGWA and LGWA$+$LISA multiband sources, respectively, with a minimum $\mathrm{S/N}_{\rm net} \simeq 11.3$.
In the EL model, the visible structures at low chirp masses reflect the adopted discrete BH seed masses combined with the inefficient growth of low-mass BHs, whereas they are largely erased in the SE scenario due to rapid gas accretion. }
\label{fig:SNR_all_scatter}
\end{figure*}

The EL and SE accretion scenarios lead to different predictions for the GW-detectable BBH population, reflecting the impact of the adopted growth prescriptions. Figure~\ref{fig:SNR_all_scatter} shows the redshift-chirp mass distribution of merging BBHs, in the EL (top) and SE (bottom) models. The columns correspond to ET (left), LGWA (middle), and LISA (right). The colored points represent detectable systems 
shaded by their S/N over 4 years of observations, while light gray points represent the full underlying (intrinsic) population in \texttt{CAT}. The median mass ratio ($\tilde{q}$) of the detectable sample is reported in each panel.

%%% EL model
In the EL model, detectable mergers span $4 < z \lesssim 14$ for all instruments and are dominated by nearly equal mass binaries ($\tilde{q} \sim 0.75$–$0.82$) at median redshift $\tilde{z}\sim 8-9$. 
The detection fraction, defined as the fraction of the observable sources that lies within the frequency domain of the instrument (including multiband sources), is predicted to be highest in the mass range probed by ET,  $\sim (10^2-10^{3.5})~\mathrm{M}_\odot$.

LISA follows with $f_{\rm LISA}\sim 30\%$ covering the range $\sim(10^5-10^8)~\mathrm{M}_\odot$, while LGWA recovers $f_{\rm LGWA}\sim 16\%$ of the detectable sample, with binaries in the intermediate mass regime, $(10^3-10^6)~\mathrm{M}_\odot$. Most detectable events, $\sim 86\%$ across all three detectors, have $8 \leq \mathrm{S/N}<50$, while only $\sim 4\%$ reach $\mathrm{S/N} \geq 100$, all within the LISA band (see Table~\ref{tab:detectability}).
%%%%%%%%%%%%%%%%%%%%%%%%%%%%%%%%%%%%%%%%%%%%%%%%%%%%%%%%%%%%%%%%%
%%%%%%%%%%%%%%%%%%%%%%%%%%%%%%%%%%%%%%%%%%%%%%%%%%%%%%%%%%%%%%%%%
%%% SE model

In the SE scenario, detectability shifts toward higher mass binaries and more asymmetric systems ($\tilde{q} \sim 0.2$–$0.4$) and extends to $z\sim 16$. ET detections are concentrated at $4<z<8$ with few $\sim 10^4 \, \rm M_\odot$ and $S/N\sim 10$ events at $z>12$. 
The lack of detectable sources at $8<z<12$ primarily reflects the rapid growth of light seeds, which shifts mergers toward higher masses and more asymmetric configurations; the predominance of unequal-mass binaries ($q \sim 0.02$–$0.08$) further reduces detectability, as their weaker GW emission falls below the ET sensitivity threshold.

ET and LGWA together would recover $\lesssim 20\%$ of the detectable population in the binary mass range $100-10^6~\mathrm{M}_\odot$, mostly with $8 \leq \mathrm{S/N} < 50$. In contrast, LISA would capture $f_{\rm LISA}\sim 80\%$ of the entire observable sample, spanning $\sim (10^{4.5}-10^{8.5})~\mathrm{M}_\odot$, with a median $\tilde{S/N} \sim 54$ and $\sim 20\%$ of the detections above $\mathrm{S/N} = 100$.  
This behavior reflects the adopted accretion prescription, in which sustained growth of the primary BH rapidly shifts mergers toward higher total masses. As a consequence, systems preferentially enter the LISA band, while the number of lower-mass binaries accessible to ET is significantly reduced. \\
Table~\ref{tab:detectability} summarizes, for each detector and accretion model over a 4-year observation period, the median mass ratio ($\tilde{q}$), median S/N ($\tilde{S/N}$), and the fractional contribution in three S/N intervals (normalized to the total detectable population) and the total detection rate ($yr^{-1}$) with $\pm 1\sigma$ scatter over ten realizations.
%%%%%%%%%%%%%%%%%%%%%%%%%%%%%%%%%%%%%%%%%%%%%%%%%%%%%%%%%%%%%%%%%
\begin{table*}[htbp]
\caption{Summary of merging BBH properties and detectability in the \texttt{CAT} model.}
\label{tab:detectability}
\centering
% ---------------------
% BLOCCO SUPERIORE
% ---------------------
\textbf{Eddington-limited model}\\[1pt]
\makebox[\textwidth][c]{
\begin{tabular}{|c|c|c|c|c|c|c|}
\hline
{\bf Detector} & {\bf $\tilde{q}$} & {\bf $\tilde{S/N}$} & {\bf f$_{8\leq S/N<50}$} & {\bf f$_{50\leq S/N<100}$} & {\bf f$_{\geq 100}$} & {\bf Detection rate (yr$^{-1}$)}\\
\hline
{ ET} & 0.7469$\pm$0.0008 & 25.45$\pm$0.33 & 0.514 $\pm$ 0.003 & 0.022 $\pm$ 0.001 & 0.001 $\pm$ 0.000  & 63.71 $\pm$ 12.00 \\
{ LGWA} & 0.8148$\pm$0.0008 & 12.39$\pm$0.11 & 0.160 $\pm$ 0.001 & 0.000$\pm$0.000 & 0.000$\pm$0.000 & 21.36$\pm$5.22 \\
{ LISA} & 0.7651$\pm$0.0003 & 43.81$\pm$0.31 & 0.182 $\pm$ 0.002 & 0.078 $\pm$ 0.001 & 0.043 $\pm$ 0.001 & 31.74$\pm$0.91 \\
\hline
\end{tabular}
}
\par\addvspace{8pt}
% ---------------------
% BLOCCO INFERIORE
% ---------------------
\textbf{Super-Eddington model}\\[1pt]
\makebox[\textwidth][c]{
\begin{tabular}{|c|c|c|c|c|c|c|}
\hline
{\bf Detector} & {\bf $\tilde{q}$} & {\bf $\tilde{S/N}$} & {\bf f$_{8\leq S/N<50}$} & {\bf f$_{50\leq S/N<100}$} & {\bf f$_{\geq 100}$} & {\bf Detection rate (yr$^{-1}$)}\\
\hline
{\bf ET} & 0.4298$\pm$0.0236 & 19.89$\pm$0.24  & 0.029 $\pm$ 0.001 &  0.000$\pm$0.000 & 0.000$\pm$0.000 & 3.78$\pm$3.33 \\
{\bf LGWA} & 0.3393$\pm$0.0020 & 13.30$\pm$0.08 & 0.174 $\pm$ 0.002 & 0.000$\pm$0.000 & 0.000$\pm$0.000 & 11.85$\pm$3.60 \\
{\bf LISA} & 0.1824$\pm$0.0006  & 53.92$\pm$0.48 & 0.413 $\pm$ 0.003 & 0.180 $\pm$ 0.003 & 0.203 $\pm$ 0.002 & 63.47$\pm$0.33 \\
\hline
\end{tabular}
}
\tablefoot{
Median mass ratio ($\tilde{q}$), median signal-to-noise ratio ($\tilde{\mathrm{S/N}}$), the fraction of detections in different $\mathrm{S/N}$ intervals, and the total detection rate are reported for the ET, LGWA, and LISA bands. Median values are given with $\pm$ 1$\sigma$ scatter over ten realizations.
}
\end{table*}
%%%%%%%%%%%%%%%%%%%%%%%%%%%%%%%%%%%%%%%%%%%

\subsection{Merger rates in the ET, LGWA, and LISA domain}

For each detector, we compute the rate of observable sources as follows \citep[e.g.,][]{Arun2009,barausse2020}:

\begin{equation}
{\rm 
\frac{\mathrm{d^2}N}{\mathrm{d}t \,\mathrm{d}z} = \frac{\mathrm{d}n}{\mathrm{d}z} 4\pi c \, \left(\frac{d_{\rm L}(z)}{1+z}\right)^2~,
}
\end{equation}

\noindent where $\mathrm{d}n/\mathrm{d}z$ is the comoving number density of detectable binaries (number of detectable binaries per unit comoving volume per unit redshift), obtained by weighting each BH merger event according to the cosmological abundance of its host halo population, summing the weighted events in redshift bins, and normalizing by the bin width and the number of merger-tree realizations (ten). As a result, the predicted GW merger and detection rates correspond to the ensemble-averaged cosmological halo population rather than the finite number of simulated merger trees. Here, $d_L(z)$ is the luminosity distance to the source. The resulting number of detections is normalized by an observation period of $t_{\rm obs}=4$ yr to obtain the expected number of detections per year.

As presented in Table~\ref{tab:detectability}, in the SE model, the total LISA detection rate is $\sim 64$ yr$^{-1}$, twice the value predicted in the EL scenario ($\sim 32$ yr$^{-1}$). In contrast, ET is strongly disfavored in the SE case, with a rate of only $\sim 4$ yr$^{-1}$ compared to $\sim 64$ yr$^{-1}$ in the EL model, a factor of $\sim 16$ difference. For LGWA, the two scenarios produce comparable detection rates, $\sim 21 \, \mathrm{yr}^{-1}$ (EL) and $\sim 12 \, \mathrm{yr}^{-1}$ (SE). 

\begin{figure}%[htbp]
\centering
\includegraphics[width=0.45\textwidth]{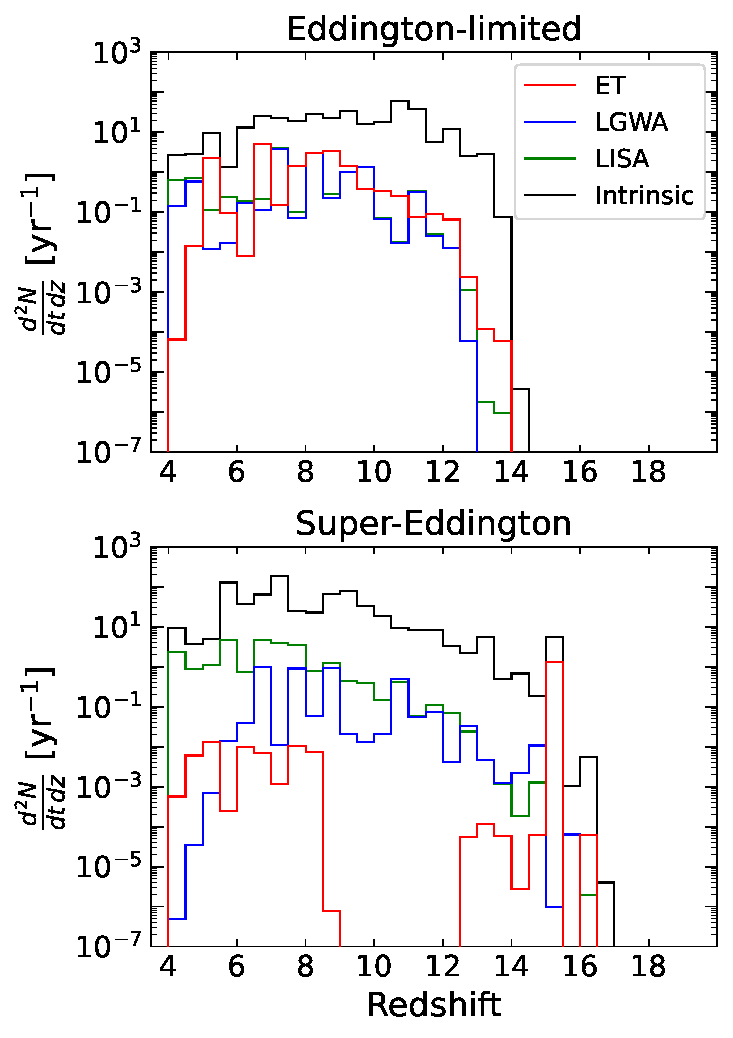} 
\caption{Detection rates of BBHs per unit time per unit redshift, $d^2N/dtdz$, for EL (top) and SE (bottom) models. Intrinsic merger rates (solid black line) are derived from the simulated population, while detectable rates for ET (red), LGWA (blue), and LISA (green) were computed assuming an S/N ratio threshold of $S/N\geq8$ calculated over a nominal observing time of 4 years, using the corresponding detector response. The resulting detection rates are normalized per year. }
\label{fig:merger_rate}
\end{figure}

Figure~\ref{fig:merger_rate} shows the intrinsic merger rate of \texttt{CAT} binaries (black histograms), together with the corresponding detection rates for ET, LGWA, and LISA in the EL (top) and SE (bottom) scenarios. The intrinsic rate is shown independently of detectability, while colored curves indicate the subset of mergers observable by each instrument, with S/N computed for a 4-year mission and normalized to give the detection rate per year. 

The two accretion prescriptions lead to qualitatively different observable rate evolutions. In the EL model, the detection rates of ET, LGWA, and LISA broadly follow the intrinsic merger rate evolution trend, with partially overlapping sensitivity domains, reflecting the coexistence of low- and high-mass systems in this scenario. The observable population is dominated by nearly equal-mass binaries, allowing substantial overlap between ground- and space-based detectors.
The mass segregation in three distinct populations reflects the adopted seeding prescription combined with Eddington-limited growth (see Fig.~\ref{fig:SNR_all_scatter}). Super-Eddington growth fills the gaps \citep[see also][]{trinca2022}, and mergers of starving low-mass binaries ($<10^{4}\, \rm M_\odot$) are confined to $4<z<8$.

In the SE regime, at $z \gtrsim 14$ mergers are dominated by intermediate-mass binaries, mainly in the ET domain. At later times, $z\lesssim 10$, the LISA detection rate increases, while the ET and LGWA rates decline as progressively fewer systems remain in their mass range. The SE model predicts a suppression of ET detections in the interval $9 \lesssim z \lesssim 12$, despite a non-negligible intrinsic merger rate. In this redshift range, mergers are typically highly asymmetric ($\tilde{q} \ll 0.1$) and more massive, which reduces the effective chirp mass in the ET band and weakens the GW signal below its sensitivity threshold, producing bimodal behavior (suppression of ET detections at $8\lesssim z\lesssim 12$). 
%%%%%%%%%%%%%%%%%%%%%%%%%%%%%%%%%%%%%%%%%%%%%%%%%%%%%%%%%%%%%%%%%
%%%%%%%%%%%%%%%%%%%%%%%%%%%%%%%%%%%%%%%%%%%
At $4\leq z\lesssim 9$ we identify a population of low-mass un-grown binaries ($<10^3 \, \rm M_\odot$) with $\tilde{q}\approx 0.3$, descending from poorly accreting ("starving") light seeds \citep[][]{valiante2021} that paired at $7<z<15$. These systems represent $\sim 1\%$ of the total merging binary population, with a comparable fraction predicted in the EL scenario over the same redshift range. With ET, we will have the unique opportunity to detect these starved binaries at high redshift, with $S/N \sim 10-20$.

Overall, the distinct redshift evolution and instrumental hierarchy predicted in the SE model, characterized by early ET sensitivity, subsequent LISA dominance, and a mid-redshift suppression of ET detections, constitute a potentially observable imprint of super-Eddington growth in the high-redshift BH population. The cosmic evolution of detection rates therefore provides a direct diagnostic of the underlying accretion physics.

%%%%%%%%%%%%%%%%%%%%%%%%%%%%%%%%%%%%%%%%%%%%%%%%%%%%%%%%%%%%%%%%%
\subsection{Gravitational wave horizons for binary black holes}
\begin{figure*}%[htbp]
\centering
\includegraphics[width=\textwidth]{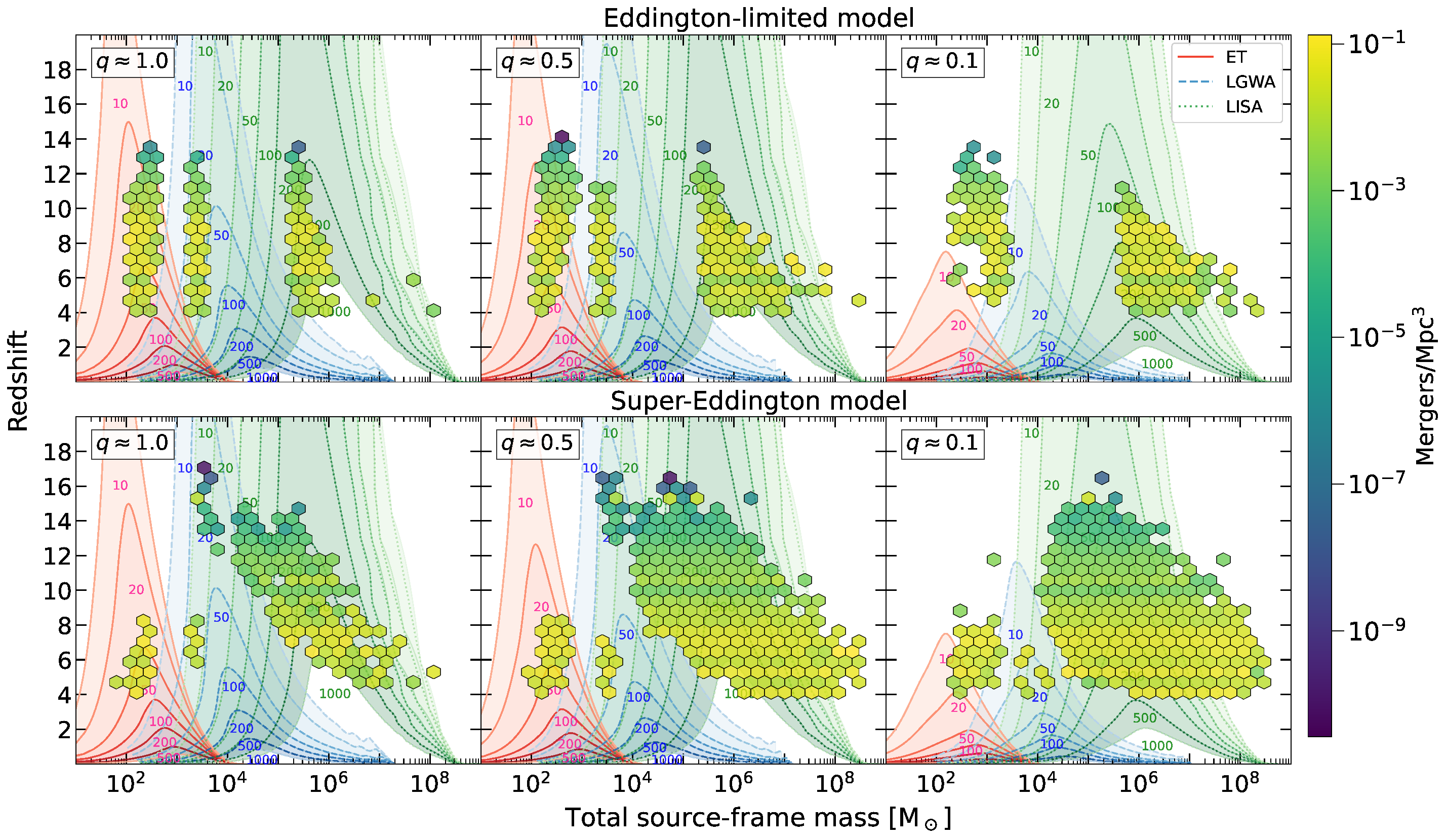} 
\caption{Signal-to-noise ratio contours for ET (red), LGWA (blue), and LISA (green) at different mass ratios: $q \approx 1$, 0.5, and 0.1. Solid, dashed, and dotted lines correspond to $S/N$ thresholds of $10$, $20$, $50$, $100$, $200$, $500$, and $1000$ for ET, LGWA, and LISA, respectively. The distribution of the BBH mergers from the EL (top) and the SE (bottom) models are shown as hexagon-binned scatter plots. Each mass-redshift cell is color-coded by the mergers number density, with selections for comparable mass ratios. 
}
\label{fig:waterfall}
\end{figure*}

To explore how different detectors probe the parameter space in binary mass and cosmic time, in Fig.~\ref{fig:waterfall} we show the redshift distribution of BBHs as a function of their total source-frame mass with overlaid curves of constant signal-to-noise ratio, $S/N=10, \, 20, \, 50, \, 100, \, 200, \, 500$, and $1000$. The curves, often called "waterfall" plots, illustrate detectability thresholds across interferometers,  ET (red), LGWA (blue), and LISA (green). 
The hexagonal bins are color-coded by the merger number density (Mpc$^{-3}$), as indicated by the color bar. The S/N curves are shown for $q=1,0.5, 0.1$, from left to right, highlighting how detectability varies between instruments as a function of binary asymmetry. The simulated binaries are selected within broader ranges ($q\geq 0.7$, $0.3\leq q<0.7$, and $0.08\leq q<0.3$) to capture the intrinsic population scatter. %and avoid artificially restricting the sample.
In the EL model, nearly equal mass binaries ($q\sim 1$) are predominantly found at $z\sim 4-8$ with typical number densities $10^{-3}-10^{-1} \, \rm Mpc^{-3}$ in the frequency domain of all three detectors. At higher redshifts, detectable binaries are rare ($<10^{-3} \, \rm Mpc^{-3}$) and still lie in the ET, LGWA, and LISA domain.
The population of moderately asymmetric systems (q$\approx$0.5) remains concentrated at $z\sim 4-12$ with typical densities around $10^{-3}-10^{-1} \, \rm Mpc^{-3}$ but with massive BBHs ($>10^7 \, \rm M_\odot$) at $z<8$ being more common than in the equal-mass regime.
As the mass ratio decreases to $q \simeq 0.1$, mergers shift toward higher total masses, reaching $\sim 10^{6}$–$10^{7}\,M_\odot$, thus lying almost exclusively within the LISA sensitivity region with number densities $\gtrsim 10^{-2}\,\mathrm{Mpc}^{-3}$ at $z\sim 5-10$. The lack of detectable sources in the ET domain reflects the higher median mass ratio, predicted for lower-mass binaries, compared to the more massive population (see Fig.~\ref{fig:SNR_all_scatter}).

The SE scenario systematically shifts the merging population toward higher masses, higher redshifts, and lower mass ratios (see also Appendix~\ref{appA:mass_ratio}. Even nearly equal-mass systems cluster at $\sim 10^{4}-10^{7}\, \rm M_\odot$ and $z \sim 6-14$ with number densities reaching $\sim 10^{-3}-10^{-1}\,\mathrm{Mpc}^{-3}$. This shift places most of the population firmly within the LISA sensitivity region, while only the lower-mass edge remains accessible to LGWA and only later ($4<z<8$) to ET. For intermediate-mass ratios, the distribution extends to $\sim 2 \times 10^3-10^8 \, \rm M_\odot$ across a wider redshift interval, $z\sim 6-16$, with densities up to $10^{-2}-10^{-1} \, \rm Mpc^{-3}$. Compared to the EL case, the population is more continuous and more extended in both mass and redshift, resulting in a clear dominance of LISA detections and a reduced contribution from ET.
The most striking difference appears in the low-mass ratio bin ($0.08<q<0.3$). The SE model predicts a substantial population of asymmetric mergers spanning $10^4-10^8 \, \rm M_\odot$ and extending to $z\sim 15-16$ with number densities reaching $10^{-2}-10^{-1} \, \rm Mpc^{-3}$ below $z\sim 10$. These systems lie almost entirely within the LISA band.

In general, the SE prescription produces a broader and more top-heavy mass distribution. The relative dominance of LISA detections and the abundance of asymmetric massive systems could be an observational signature that distinguishes SE growth from EL evolution. We recall that in this work we specifically track the merger histories of cosmic binaries (see Sect.~\ref{sec:cosmic_BBH}), not including the population of stellar-origin BBHs evolving in situ.
%%%%%%%%%%%%%%%%%%%%%%%%%%%%%%%%%%%%%%%%%%%%%%%%%%%%%%%%%%%%%%%%%%%%%%%
\subsection{Multiband gravitational wave sources}
\label{sec:multiband}
We explore the population of potential multiband sources within the conceptual framework introduced in Sect.~\ref{sec:cosmic_BBH}. As an illustrative scenario, we assume that LISA, LGWA, and ET operate simultaneously over a common 4-year observation window, with merger times of all systems randomly distributed within this interval.

\begin{figure*}%[htbp]
\centering
\includegraphics[width=\textwidth]{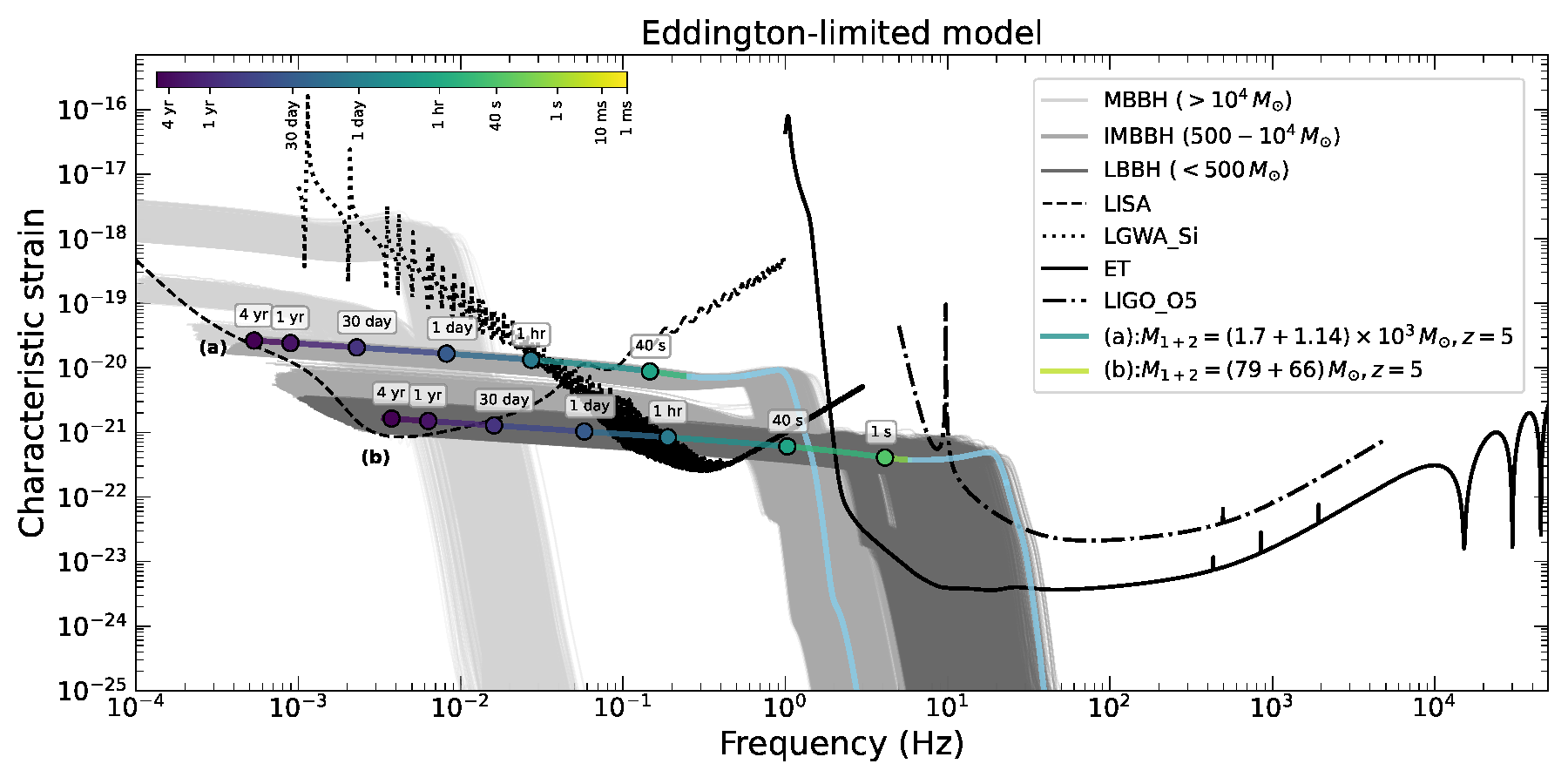} 
\caption{Characteristic strain of BBH mergers in the EL scenario over 4-year observations and assuming an optimal (face-on) orientation. Sensitivity curves for LIGO, ET, LGWA, and LISA (including instrumental noise and Galactic foreground) are shown for reference. Gray tracks represent the full population categorized by total mass: massive BBHs (>$10^4$ M$_{\odot}$, faint gray), intermediate-mass BBHs (500–$10^4$ M$_{\odot}$, medium gray), and low-mass BBHs (<500 M$_{\odot}$, dark gray). The solid colored lines (a) and (b) indicate representative intermediate-mass and stellar-mass BBHs, with markers showing the frequency at selected times before merger, from 4 yr down to 1 ms. System (b) could illustrate a multiband detectability, as LISA observes its early inspiral nearly 4 years before merger, LGWA tracks the late inspiral approximately 1 hour pre-merger, and ET captures the final few seconds before merger. The merger and ring down phases are shown in light blue. The horizontal color bar encodes the time to coalescence. 
}
\label{fig:char_strain_EL}
\end{figure*}

In Fig.~\ref{fig:char_strain_EL}, we present the characteristic strain $h_c(f)$ as a function of frequency for the intrinsic population of merging BBHs, with $0.001< q \lesssim 1$, in the EL model. 
This is computed by combining the two GW polarizations $h_+$ and $h_\times$ and accounting for the time spent by the signal in a given frequency band:

\begin{equation}
    h_c(f) = 2f \sqrt{ |h_{+}(f)|^2 + |h_{\times}(f)|^2 }.
\label{eq:characteristic_strain}
\end{equation}
Synthetic strain spectra are compared with the sensitivity curves of LISA, LGWA, ET, and LIGO (O5 run).

In Fig.~\ref{fig:char_strain_EL}, the binaries are classified by total mass into massive BBHs (MBBHs; $>10^{4}\,M_\odot$, light gray), intermediate-mass BBHs (IMBBHs; $500$--$10^{4}\,M_\odot$, medium gray) and low-mass BBHs (LBBHs; $<500\,M_\odot$, dark gray). The colored solid tracks (a) and (b) highlight two representative candidates in the intermediate- and low-mass regimes, respectively. Circles along each track mark the GW frequency at selected times before merger, ranging from $t_{\mathrm{obs}} = 4$ yr down to the time when the system reaches the observed ISCO frequency. The post-ISCO regime beyond the inspiral phase is highlighted in light blue. We computed the time to coalescence as \citep{sesana2016}: 
\begin{equation}
    t_r = \frac{5c^{5}}{256 \pi^{8/3}} (G \mathcal{M}_r)^{-5/3} f_r^{-8/3}\, ,
\end{equation}

where the subscript $_r$ refers to the quantities measured in the rest frame of the source. As expected, MBBHs dominate at frequencies $\sim 10^{-4}$–$10^{-1}$ Hz (LISA band) and $\sim$ $10^{-3}-0.3$ Hz (LGWA band), while IMBBHs and LBBHs are also detectable at higher frequencies ($f \gtrsim 10^{-1}$ Hz). In particular, IMBBHs could overlap with LISA, LGWA, and ET sensitivity bands, while LBBH mergers are detectable only by ground-based detectors.

Source (a) is an IMBBH merging at redshift $z \sim 5$, with primary mass $M_1 \approx 1700\, \mathrm{M}_\odot$ and $q \approx 0.7$. The timeline at the top of the panel shows the frequency evolution of this system during the final 4 years before coalescence. 
Systems like this are promising multiband candidates. The early inspiral falls in the LISA band (e.g., $f\sim 0.53$ mHz about 4 years before merger), the late inspiral enters the LGWA sensitivity window ($\sim 0.27$ dHz $\sim 1\, \rm hour$ before merger) whereas the merger–ringdown phase is entirely inaccessible to ground-based detectors.

Source (b) is a LBBH at $z\sim 5$, with primary mass $M_1 \approx 80 \, \rm M_\odot$ and $q\sim 0.8$. Such binaries are accessible to all three future observatories. LISA would observe the inspiral phase about 4 years before the merger at $\sim 3.8$ mHz. Approximately one hour before merger, the late inspiral phase enters the LGWA band at $\sim 0.2$ Hz, and just a few seconds before merger, it reaches the ET band at $\sim 2$ Hz.

Finally, we computed the median fraction of potential multiband events, detectable by two or three instruments, relative to the total detectable population in each model.
We define multiband events as systems detectable by at least two observatories and compute their fraction relative to the total detectable population in each model. Since we require $\mathrm{S/N} \geq 8$ in each individual detector, multiband detections correspond to a higher effective network signal-to-noise ratio 
$\mathrm{S/N}_{\rm net} = \sqrt{\sum_i (\mathrm{S/N}_i)^2}$
where the sum runs over the detectors composing the network.
This implies a minimum $\mathrm{S/N}_{\rm net} \simeq 13.9$ and $\simeq 11.3$ for three- and two-detector networks, respectively.
In both accretion scenarios, network performance is determined primarily by the overlap between the underlying BH mass distribution and the detectors’ sensitivity bands. As a result, networks including ET$+$LISA provide small multiband overlap, and the full ET$+$LGWA$+$LISA network captures only a tiny fraction of sources, $f_{ET+LGWA+LISA}<0.01\%$ ($0.002\%$) in the EL (SE) scenario. In the EL model, although low-mass binaries dominate the merging population, favoring ET sensitivity, ET-based networks capture very few multiband candidates. The largest fraction of sources are candidates for multiband in the LGWA$+$LISA network $f_{LGWA+LISA}\sim 24\%$ significantly higher than ET$+$LGWA ($f_{ET+LGWA}\sim 0.016\%$). The multiband candidates in ET-based networks have total masses of $\sim (2-3)\times10^3\, \rm M_\odot$ and mass ratio $q\sim 0.6 - 0.8$ ($\tilde{q}\sim 0.7$) at $z\sim 4-5$, while in the LGWA$+$LISA network, more massive systems are selected, $\sim 10^5-10^6 \, \rm M_\odot$ with $q \sim 0.3-1$ ($\tilde{q}\sim 0.8$) at $z\sim 4 - 13$.

In the SE model, rapid mass growth shifts multiband candidates toward higher masses and lower mass-ratios, making space-based detectors the primary drivers of multiband sensitivity. The predicted sources in the LGWA$+$LISA network ($f_{LGWA+LISA}\sim 11\%$) have masses of $\sim 10^4$–$10^6\, \rm M_\odot$ and $0.03<q<1$ ($\tilde{q}\sim 0.3$) at $z\sim 4.4-16$, while possible targets of ET-based networks ($f_{ET+LGWA}\sim 0.005\%$) are restricted to a narrower mass interval $\sim (3-5)\times10^3 \, \rm M_\odot$ at $z\sim 4.4-6$ and with a mass ratio range of $q\sim 0.3 - 0.5$ ($\tilde{q}\sim 0.4$). In Fig.~\ref{fig:SNR_all_scatter}, we show the multiband source overlap regions for ET$+$LGWA (circular markers) and LGWA$+$LISA (dashed lines), with the combined S/N$>$11.3 in each case. 

%%%%%%%%%%%%%%%%%%%%%%%%%%%%%%%%%%%%%%%%%%%%%%%%%
%%%%%%%%%%%%%%%%%%%%%%%%%%%%%%%%%%%%%%%%%%%

%%%%%%%%%%%%%%%%%%%%%%%%%%%%%%%%%%%%%%%%%%%%%%%%%%%%%%%%%%%%%%%%%

\section{Discussion}
\label{sec:discussion}

Our semi-analytical model, \texttt{CAT}, provides a computationally efficient framework to statistically investigate the assembly of massive BHs and the formation of BBHs, together with their GW signatures, during the first $\sim1.5$ Gyr of cosmic history. Although the model does not explicitly resolve the spatial structure of galaxies or the orbital evolution of individual BHs, it enables extensive parameter studies over cosmological volumes that remain beyond the reach of current high-resolution hydrodynamical simulations.

\noindent 
\subsection{Model assumptions and physical uncertainties}
 
The simplified prescriptions adopted for BH accretion reflect the current uncertainties in modeling gas inflows across the wide range of spatial scales involved. The EL prescription likely underestimates the growth of light- and medium-weight BH seeds, whereas the merger-driven SE scenario may instead overestimate the abundance of the most massive systems detectable by LISA.
Similarly, assuming gas accretion only onto the primary BH maximizes the mass-ratio asymmetry and should therefore be regarded as a conservative limiting case. Allowing for some accretion onto the secondary BH during the pairing phase could reduce this asymmetry, as tidal perturbations can drive gas toward the secondary nucleus \citep[e.g.,][]{capelo2015}. However, the efficiency of secondary accretion before binary formation remains poorly constrained. The secondary BH is expected to spend extended periods of the pairing phase at kpc-to-hundreds-of pc separations from the galaxy center, outside the dense circumnuclear gas reservoir \citep[e.g.,][]{Fiacconi2013, Roskar2015, SouzaLima2017}. Nevertheless, accretion from circumnuclear/circumbinary environments during the subsequent hardening phase \citep[e.g.,][]{Duffell2020, Dittmann2021, Siwek2023, Franchini2024}, may modify the binary mass ratio before coalescence.

To isolate the impact of different seeding channels, medium-weight and heavy BHs are currently assigned fixed birth masses. Although this facilitates the interpretation of the resulting BBH populations, radiation-hydrodynamical simulations predict broader seed mass distributions that depend on the thermodynamical and radiative properties of the host halo \citep[e.g.,][]{ReganVolonteri24,prole2024,McCaffrey2025,chon2025,Chon2026}. Implementing physically motivated seed-mass distributions is therefore a natural extension of \texttt{CAT} and is expected to smooth the discrete features visible in the EL model while preserving the main evolutionary trends.

The delay between galaxy and BH mergers is computed assuming dynamical-friction-driven orbital decay, which is expected to dominate the early stages of BH pairing \citep[e.g.,][]{mayer2007,capelo2015,biava2019,Li2022}. The resulting delays are broadly consistent with previous semi-analytical models and dedicated simulations \citep[e.g.,][]{volonteri2003, volonteri2020, barausse2020, Chen2022, Li2022, Chakraborty2023, Izquierdo2024, Langen2025}. Nevertheless, the pairing phase remains one of the largest sources of uncertainty in predicting massive BH merger rates \citep[e.g.,][]{izquierdoVillalba2026,Singh2026}. Additional processes neglected here may halt, delay, or accelerate BH coalescences, including global or bar-driven torques \citep[e.g.,][]{bortolas2020}, stellar or gas-driven hardening, and the final GW-driven inspiral \citep[][]{begelman1980,colpi2014}, which are generally expected to proceed rapidly once a bound binary has formed \citep[e.g.,][]{Dotti2007,Sesana2015,bortolas2016,bortolas2018,ArcaSedda2019,biava2019,souzaLima2020,franchini2021,franchini2022}, post-merger gravitational recoil \citep[e.g.,][]{merritt2004,campanelli2007kick,kelley2017,volonteri2010recoil,Ricarte2025}, and triple/multiple BH interactions \citep[][]{bonetti2018a,bonetti2018b,bonetti2019,valiante2021,Sayeb2024}. Future versions of \texttt{CAT} will incorporate these processes to provide a more complete description of BBH evolution.

Despite these simplifications, \texttt{CAT} provides a flexible framework for exploring the statistical impact of BH seeding, growth, and merger delays on the high-redshift BBH population, while maintaining the computational efficiency required for large cosmological parameter studies. The inclusion of additional physical processes, such as more realistic gas inflows, angular-momentum transport, AGN feedback, and BH spin evolution, represents a natural direction for future developments \citep[e.g.,][]{negri2017,waters2020,Ranjbar2025bondi,Tripathi2025bondi,Zana2025}.

\subsection{Comparison with previous studies}

The merger and detection rates predicted by \texttt{CAT} lie within the broad range of current theoretical studies, although they generally exceed those obtained from large-volume cosmological simulations focused on $z\lesssim4-6$ \citep[e.g.,][]{kelley2017,tremmel2018,degraf2020, volonteri2020,katz2020}. This mainly reflects the scientific focus of \texttt{CAT} on the first $\sim 1.5$ Gyr of cosmic history, when BH seeding and galaxy mergers are the most efficient, enhancing the formation of merging BBHs.
Interestingly, the LISA detection rates predicted in our SE model at $5<z<15$ are broadly consistent with those inferred from the high-resolution Renaissance simulations, in which BH seeding and dynamical evolution are followed in post-processing \citep[][]{McCaffrey2025}.

Among semi-analytical models, our predicted LISA detection rates lie at the lower end of light-seed scenarios, while remaining above most heavy-seed models \citep[e.g.,][]{sesana2007,klein2016,ricarte2018b, bonetti2019, dayal2019, Ricarte2025}. In particular, the SE model predicts rates comparable to those obtained by \citet{dayal2019}, whereas the EL model yields systematically lower values. Overall, the predicted detection rates remain consistent with the broad range of current theoretical expectations.

A recent systematic comparison of state-of-the-art massive BH evolution models by \citet{izquierdoVillalba2026} showed that substantial differences in merger and detection rates persist even after adopting an identical prescription for the BH pairing timescale. This highlights that the dominant model-to-model scatter originates primarily from differences in BH seeding, gas accretion, feedback, galaxy evolution, and halo assembly histories, rather than from the treatment of the pairing phase alone. 
Different seeding models determine the abundance and initial mass spectrum of BHs available to form binaries, while gas accretion, stellar and AGN feedback regulate their subsequent growth. Likewise, differences in star formation, chemical enrichment, halo assembly histories, halo mass resolution, and the construction of dark matter merger trees affect both the abundance of BH seeds and the timing of galaxy mergers. Consequently, by the time the pairing phase is reached, each model has already produced a distinct BBH population in terms of masses, mass ratios, host-galaxy properties, and merger-redshift distribution.

The diversity of current predictions reflects the present uncertainties in modeling the BH/galaxy formation, while simultaneously defining the range of physically plausible evolutionary pathways. Exploring this diversity is essential to fully exploit the diagnostic power of future GW observations. 
Within this landscape, the present implementation of \texttt{CAT} introduces two complementary physical developments: (i) a third BH-seeding channel producing medium-weight seeds and (ii) a treatment of finite dynamical-friction delays between galaxy mergers and BH mergers. Together, these provide a self-consistent framework to investigate how BH seeding, growth, and delayed mergers shape the high-redshift BBH population and its GW signatures during the first $\sim 1.5$ Gyr of cosmic history.

Finally, we emphasize that the present work focuses on BBHs forming in the aftermath of galaxy encounters. Additional stellar-origin BBH populations, produced through dynamical interactions in young or globular star clusters, isolated binary evolution in galactic fields or multiple Pop~III star remnants \citep[e.g.,][]{Mapelli2021,Mapelli2022,Santoliquido2020, Hirano2018,sugimura2020}, are expected to dominate the low-mass regime accessible to ET. Although based on different formation channels, the IMBH merger rates predicted by \texttt{CAT} are broadly consistent with the lower-to-intermediate range of these stellar-origin IMBH populations \citep[e.g.,][]{Branchesi2023, Liu2024, Davari2024}.

%%%%%%%%%%%%%%%%%%%%%%%%%%%%%%%%%%%%%%%%%%%%%%%%%%%%%%%%%%%%%%
\section{Conclusions}
\label{sec:conclusions}

In this work, we investigated the formation, evolution, and gravitational wave signatures of BBHs from $z=24$ to $z=4$ using the semi-analytical model \texttt{CAT}. By comparing Eddington-limited and super-Eddington accretion models, we explored how BH seeding, growth, and dynamical delays shape the population of merging binaries detectable by LISA, LGWA, and ET. We used \texttt{GWFish} to compute the S/N for \texttt{CAT} BBHs to assess their detectability in the three detector frequency ranges, imposing a detection threshold of S/N $>8$. Assuming a merger delay time set by the dynamical friction timescale, we find that the distributions of the BBH mass and mass ratio critically depend on the accretion regime. As a consequence, the detection rates of potential GW sources in our synthetic catalog are highly sensitive to the accretion prescription, implying that joint observations across multiple GW detectors will represent a powerful probe to discriminate among different accretion mechanisms governing the early growth of the first MBHs. We summarize our findings as follows: 
\begin{itemize}
    \item About $\sim20\%$ of the BH pairs that form in the aftermath of galaxy major mergers eventually form a bound binary system and merge, by $z=4$, independently of the accretion model.

    \item The EL scenario places the majority of detectable mergers ($\sim 50\%$) in the ET mass/frequency domain, whereas the SE model shifts the detectable population toward higher masses, with $\sim 80\%$ of events falling within the LISA band.
    
    \item The EL scenario favors nearly equal mass binaries, mostly detectable by ET and LGWA up to $z\sim8$ and $12$, respectively. In contrast, the SE model predicts a detectable population of asymmetric binaries ($q \approx 0.1$) for LGWA and LISA up to $z\sim 14$.
    
    \item SE growth doubles the detection rate in the LISA domain ($\sim 64$ yr$^{-1}$) compared to the EL regime ($\sim 32$ yr$^{-1}$), while strongly reducing the rate predicted in the ET sky to $\sim 4 \, \rm yr^{-1}$. The accretion scenarios only mildly affect the detection rates in the LGWA domain.
    
    \item Characteristic strain analysis suggests that intermediate-mass systems are promising multiband candidates, especially in the EL scenario. In both models, higher masses favor space-based combinations, with LGWA$+$LISA yielding the highest fraction, approximately $\sim 24 \%$ in EL and $\sim 11\%$ in SE. 

\end{itemize}

\noindent The complementarity between future GW observatories highlights the importance of multiband observations, where and when possible, and detector synergy as a key strategy to unveil the origin, growth, and assembly of BHs in the early Universe.

%%%%%%%%%%%%%%%%%%%%%%%%%%%%%%%%%%%%%%%%%%%%%%%%%%%%%%%%%%%%%%
\begin{acknowledgements}
    We thank Marta Volonteri, Alberto Mangiagli and Jan Harms for helpful discussions.
      ND, AT, RV, and AL acknowledge support from PRIN MUR "2022935STW" funded by European Union-Next Generation EU, Missione 4 Componente 2 CUP C53D23000950006 and from the Bando Ricerca Fondamentale INAF 2023, Theory Grant "Theoretical models for Black Holes Archaeology" and Mini-grant "Cosmic Archaeology with the first black hole seeds". RV acknowledges financial support from the ASI-INAF agreement n. 2024-36-HH.1-2025. RS acknowledges support from the PRIN 2022 MUR project 2022CB3PJ3—First Light And Galaxy aSsembly (FLAGS) funded by the European Union—Next Generation EU, from EU-Recovery Fund PNRR - National Centre for HPC, Big Data and Quantum Computing, and from the INFN TEONGRAV initiative. AS is supported by the European Union’s H2020 ERC Advanced Grant ``PINGU'' (Grant Agreement: 101142079).
\end{acknowledgements}

%%%%%%%%%%%%%%%%%%%%%%%%%%%%%%%%%%%%%%%%%%%%%%%%%%
%%%%%%%%%%%%%%%%%%%%%%%%%%%%%%%%%%%%%%%%%%%%%%%%%%
%%%%%%%%%%%%%%%%%%%% REFERENCES %%%%%%%%%%%%%%%%%%

\bibliographystyle{aa} % style aa.bst
\bibliography{reference} 
%%%%%%%%%%%%%%%%%%%%%%%%%%%%%%%%%%%%%%%%%%%%%%%%%%

\begin{appendix}
\section{Mass ratios distribution of cosmic BH binaries}
\label{appA:mass_ratio}

\vspace{0.5cm}

\begin{figure}[b]
\centering
\includegraphics[width=0.728\linewidth]{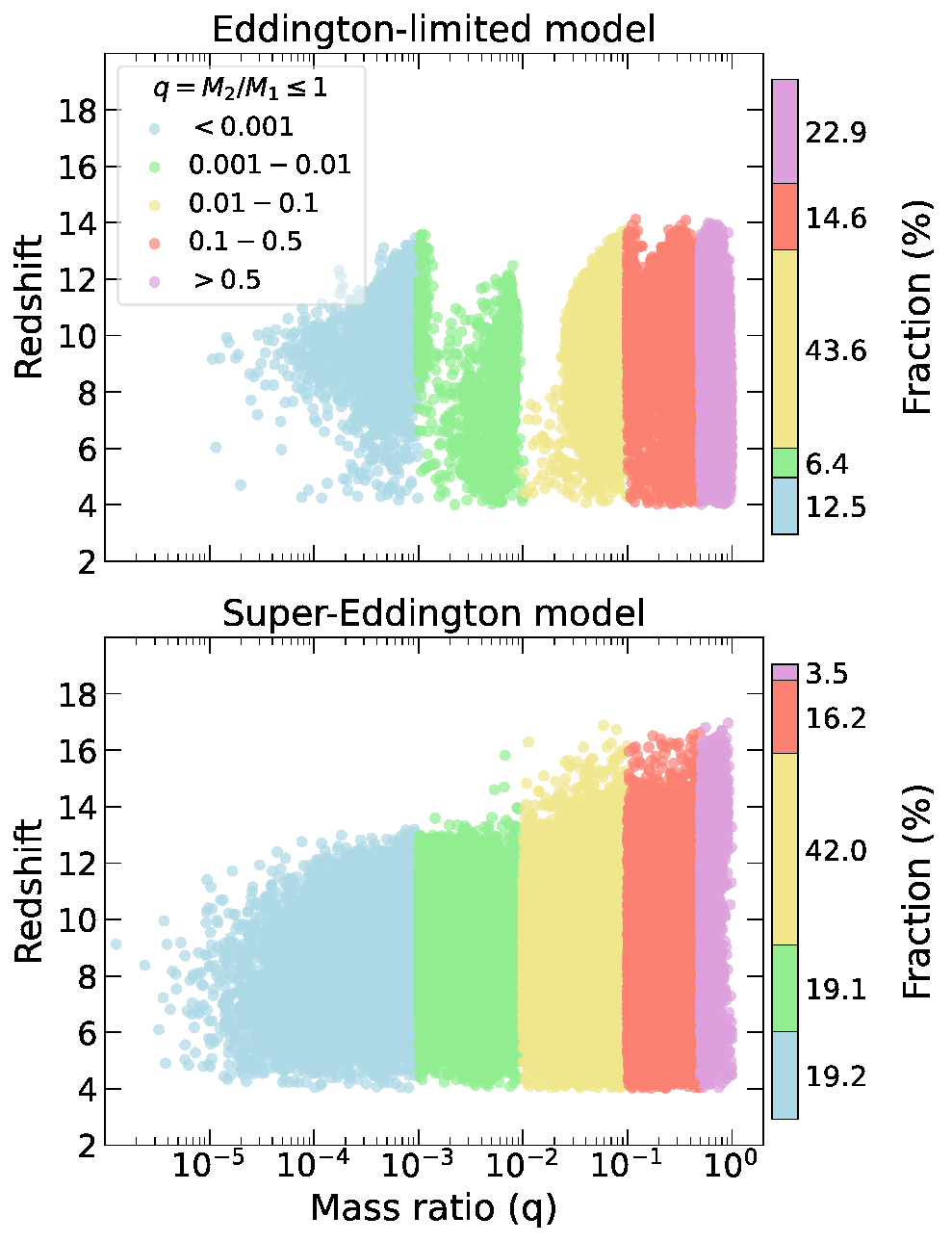} 
\caption{Distribution of mass ratios ($q=M_2/M_1$) for merging binaries in the EL (top) and SE (bottom) models, normalized to the total population of binary systems. Colors depict different $q$ intervals: $q<10^{-3}$ (blue), $10^{-3}<q<10^{-2}$ (green), $10^{-2}<q<0.1$ (yellow), $0.1<q<0.5$ (red), and $q>0.5$ (violet). The right-hand bar indicates the fractional contribution to each interval. The mass ratio decreases with redshift, as a result of differential accretion of the primary and secondary BHs, particularly in the SE model.
}
\label{fig:massRatios}
\end{figure}

In Fig.~\ref{fig:massRatios}, we show the distribution of our cosmic BBHs mass ratio, $q \equiv M_2/M_1 \leq 1$, versus the redshift for the EL and SE scenarios. As in the cosmic BBH mass distribution, the accretion mode also shapes the mass ratio distribution. We categorized the BBH mass ratios into five intervals: $q < 0.001$, $0.001 < q < 0.01$, $0.01 < q < 0.1$, $0.1 < q < 0.5$, and $q > 0.5$. Binaries with $q < 0.001$ are excluded in our estimations (see Sect.~\ref{sec:cosmic_BBH}). 

In the lowest interval ($0.001-0.01$), the SE model predicts a notably higher fraction ($19.2\%$) than the EL model ($6.4\%$). About $40\%$ of mergers have $0.01<q<0.1$ in both models and about $40\%$ of the binaries have $q>0.1$ in the EL model. In the SE scenario, the latter fraction drops below $20\%$ as the primary BH grows efficiently during the pairing phase. 

In our models, during the pairing phase, gas accretion is allowed only onto the most massive BH at the center of the merged galaxy. At the onset of this stage, binaries typically have high mass ratios, with more than $70\%$ of systems characterized by $q>0.5$ in both the EL and SE scenarios. As the primary BH continues to accrete while the secondary remains starved, the mass ratio progressively decreases. By the time of coalescence, the fraction of binaries with $q>0.5$ drops below $23\%$ (EL) and $4\%$ (SE). Although the SE model predicts a smaller relative fraction of high-$q$ systems, it yields a larger overall number of mergers (see Sect.~\ref{sec:detectability}), so the absolute merger rate of binaries with $q>0.5$ may remain comparable between the two scenarios.
%%%%%%%%%%%%%%%%%%%%%%%%%%%%%%%%%%%%%%%%%%%%%%%%%
%%%%%%%%%%%%%%%%%%%%%%%%%%%%%%%%%%%%%%%%%%%%%%%%%%%%%%%%%%%

\section{Validation of the updated \texttt{CAT} model}
\label{appB:recalibration}

Following the implementation of the delayed BH merger prescription, the AGN feedback efficiency was recalibrated to preserve the agreement between the \texttt{CAT} predictions and the observational constraints adopted in \citealt{trinca2022}. This Appendix shows that the updated model continues to reproduce the main calibration benchmarks.

Importantly, none of the GW-related quantities presented in this work was used in the recalibration procedure. Therefore, BBH merger rates, mass-ratio distributions, and GW detection statistics remain genuine predictions of the updated \texttt{CAT} model rather than quantities tuned to reproduce GW observables.

\begin{figure}
\centering
\includegraphics[width=0.85\linewidth]{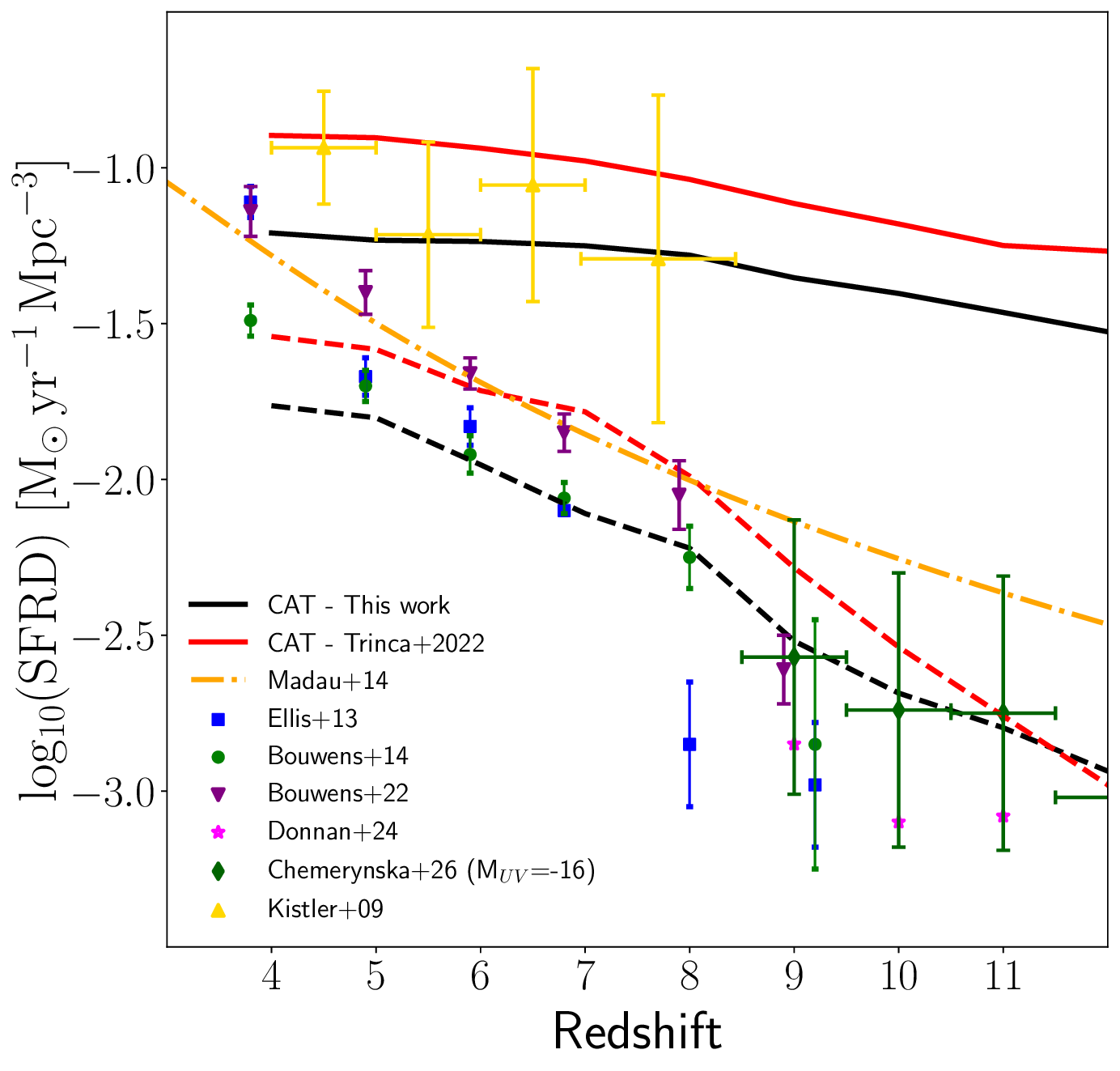}
\caption{Redshift evolution of the comoving star formation rate density (SFRD). The revised model ($\epsilon_{w,\rm AGN}=10^{-3}$; black) is compared with the original \texttt{CAT} model \citep[][red]{trinca2022}. Solid curves show the total galaxy population, while dashed curves correspond to galaxies with $M_{\rm UV}\leq -16.5$. Observational constraints, from \cite{ellis2013}, \cite{bouwens2014}, \cite{Bouwens2022B}, \cite{Donnan2024}, \cite{Chemerynska2026}, \cite{kistler2009}, \cite{madau2014b}, are overplotted. 
}
\label{fig:SFR}
\end{figure}

Figure~\ref{fig:SFR} shows the comoving star formation rate density (SFRD) as a function of redshift for the EL model at $4 < z < 8$. Solid lines show the predictions of the updated model (black) compared to the original \texttt{CAT} calibration of \citet[][red]{trinca2022}, together with the available observational constraints. The dashed lines show the corresponding UV-selected galaxy population down to the observed faint limit of $M_{\rm UV}\leq -16.5$ (black) and $M_{\rm UV}\leq -17.7$ (red). The latter corresponds to the threshold adopted in \citet{trinca2022}. The selection of a fainter limit of $M_{\rm UV}$ reflects the improved observational constraints on the faint-end of the UV galaxy population enabled by recent JWST observations. Recent results from \citet{Chemerynska2026} indicate that galaxies fainter than $M_{\rm UV} \sim -16$ contribute more than $50\%$ of the total cosmic SFRD at $z \sim 12$, highlighting the importance of accounting for the faint galaxy population when measuring the cosmic SFRD at high redshifts. Furthermore, \citet{Hainline2026} presents significant galaxy samples down to $M_{\rm UV} \sim -16$, confirming that JWST now routinely probes the faint-end population that was previously inaccessible. Our adopted limit of $M_{\rm UV} \leq -16.5$ is therefore chosen to capture the contribution of faint galaxies to the cosmic SFRD, while maintaining consistency with recent observational data.

As in the original \texttt{CAT} model, the offset between the total and UV-selected populations highlights the significant contribution of currently undetected faint galaxies to the SFRD. Compared to the original calibration, the updated model predicts systematically lower SFRD over the explored redshift range. These differences arise from the combined effect of introducing delayed BH mergers and the subsequent recalibration of the AGN wind efficiency required to preserve the agreement with the observed quasar population. Together, these modifications alter the interplay between BH growth and galaxy evolution. The weaker AGN feedback reduces the amount of gas expelled from galaxies, while the delayed-merger prescription modifies the BH assembly history across cosmic time. As a consequence, a larger fraction of the available cold gas is channeled into BH accretion, favoring BH growth over star formation compared to the original calibration. Despite these changes, the predicted SFRD remains in reasonable agreement with the available observational constraints. 

%%%%%%%%%%%%%%%%%%%%%%%%%%%%%%%%%%%%%%%%%%%%%%%%%
%%%%%%%%%%%%%%%%%%%%%%%%%%%%%%%%%%%%%%%%%%%%%%%%%%%%%%%%%%%

\begin{figure}
\centering
\includegraphics[width=\linewidth]{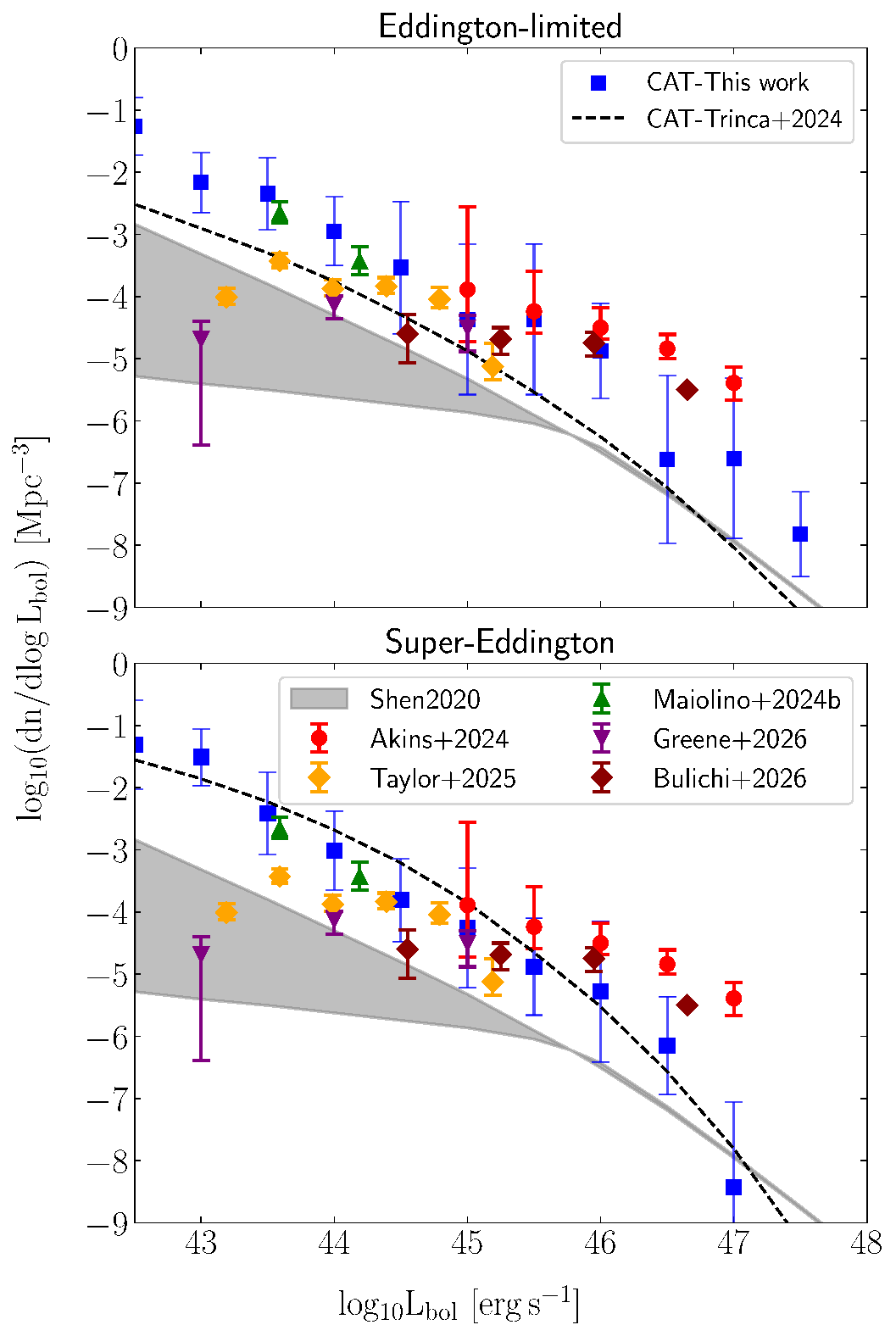} 
\caption{AGN bolometric luminosity function predicted by \texttt{CAT} at $5<z<7$ for the EL (top) and SE (bottom) accretion models. Blue squares show the revised model, while the dashed black lines correspond to the original \texttt{CAT} best fit \citep{trinca2022,trinca2024superEdd}. The gray shaded region shows the observed quasar bolometric luminosity function at $z\approx6$ from \citet{shen2020}. Recent observational results from \citet{Akins2024,Maiolino2024bhs,Taylor2025,Greene2026,Bulichi2026} are included for comparison. 
}
\label{fig:Lbol}
\end{figure}

Figure~\ref{fig:Lbol} illustrates the impact of these modifications on the AGN population, showing the bolometric luminosity functions predicted by the updated EL (top) and SE (bottom) models (blue squares), compared to the original \texttt{CAT} calibration (black dashed lines; \citealt{trinca2022,trinca2024superEdd}) and the observed quasar bolometric luminosity function from available observational data. The combined effect of the delayed-merger prescription and the adopted recalibration differs between the two accretion models. In the SE model, the revised AGN wind efficiency largely compensates for the changes introduced by delayed BH mergers, leaving the predicted bolometric luminosity function almost unchanged with respect to the original calibration while preserving the agreement with the observational data. In contrast, the EL model predicts a systematically larger AGN number density over most of the luminosity range. This reflects the stronger dependence of the EL accretion on the available cold gas reservoir, whereas its impact is less significant in the SE accretion because the latter is more episodic, and each accretion episode is limited by the gas mass in the halo. As a result, the combined effect of a weaker AGN feedback and the modified BH assembly allows BHs to grow more efficiently in the EL model, increasing the AGN abundance. This interpretation is consistent with the lower SFRD shown in Fig.~\ref{fig:SFR}, indicating that the updated model favors BH growth over star formation compared to the original calibration. Nevertheless, the predicted luminosity function remains in good agreement with the available observational constraints.
%%%%%%%%%%%%%%%%%%%%%%%%%%%%%%%%%%%%%%%%%%%%%%%%%
%%%%%%%%%%%%%%%%%%%%%%%%%%%%%%%%%%%%%%%%%%%%%%%%%%%%%%%%%%%

\section{Systems merging below $z = 4$} 
\label{appC:lower_z}

%%%%%%%%%%%%%%%%%%%%%%%%%%%%%%%%%%%%%%%%%%%%%%%%%
\begin{figure}
\centering
\includegraphics[width=0.5\textwidth]{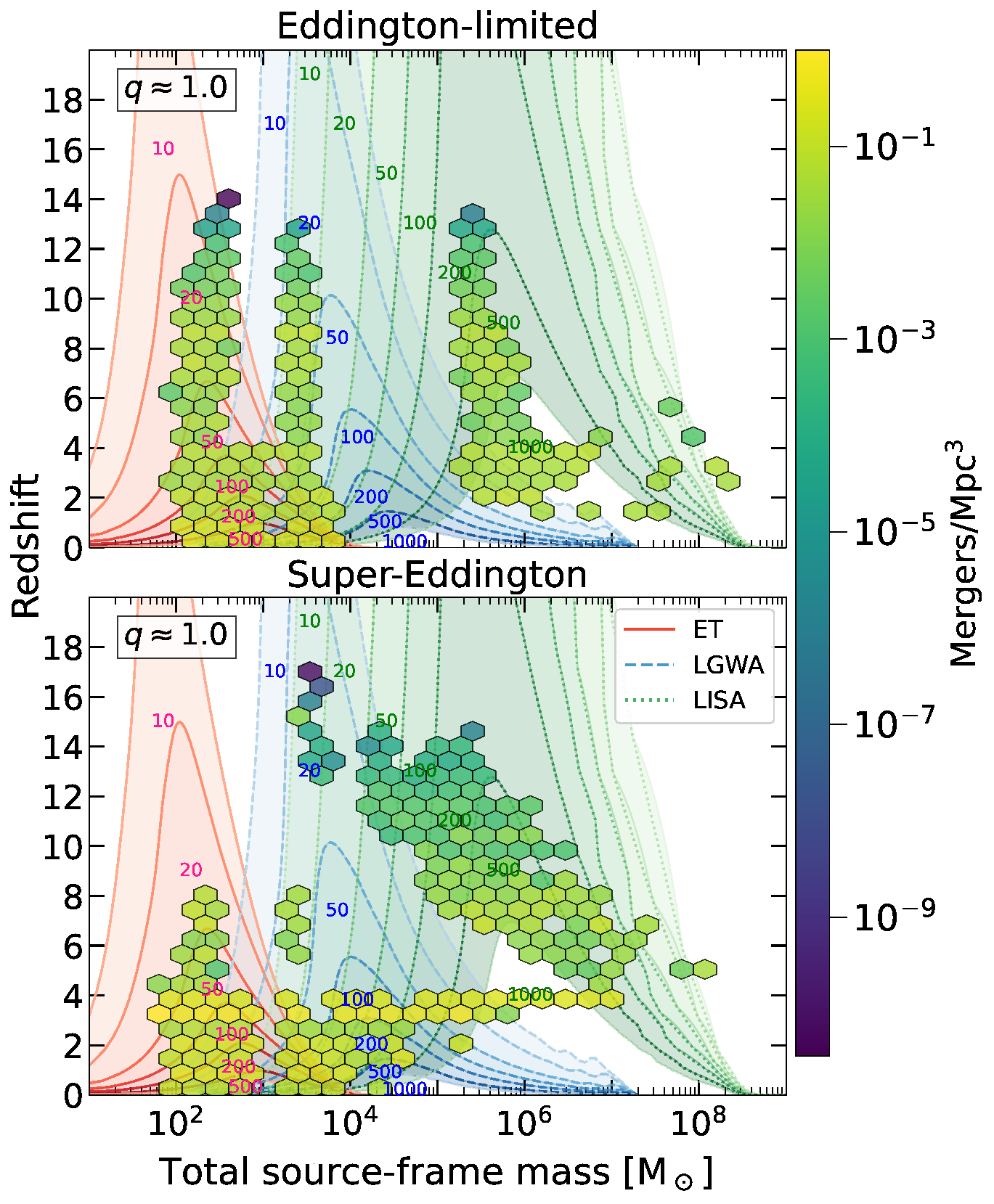} 
\caption{Same as Fig.~\ref{fig:waterfall}, but extended to include sources with $z<4$ and considering only binaries with $q \approx 1$.}
\label{fig:waterfall_lower_z}
\end{figure} 
%%%%%%%%%%%%%%%%%%%%%%%%%%%%%%%%%%%%%%%%%%%%%%%%%%%%%%%%%%%

The current version of \texttt{CAT} follows the formation and evolution of galaxies and BBHs over the redshift range $z \geq 4$, corresponding to the first $\sim 1.5$ Gyr of cosmic history. Consequently, some BBHs are still in the pairing phase when the model reaches its lower-redshift boundary, because their dynamical-friction timescales extend beyond cosmic time at $z=4$. To estimate the contribution of these systems to the lower-redshift merger population, we performed an additional post-processing analysis. 
We selected BH pairs that are still undergoing the pairing phase at $z=4$ and whose evolution has not been interrupted by subsequent galaxy mergers before the end of the simulation. We then propagated their predicted dynamical friction delay times in \texttt{CAT} beyond $z=4$ to estimate the redshift at which they would coalesce. In this extrapolation, the BBH and host-galaxy properties are kept fixed after $z=4$, so that subsequent evolution is treated passively and does not include further BH accretion, galaxy evolution, or additional interactions. The resulting coalescence redshifts should therefore be interpreted as approximate estimates of the delayed-merger population rather than fully self-consistent predictions.

For the $z<4$ population of merging binaries we computed the S/N, merger, and detection rates in the ET, LGWA, and LISA frequency bands adopting the same approach as described in Sects.~\ref{sec:cosmic_BBH} and ~\ref{sec:detectability}. For the EL model, the detectable ET and LGWA populations remain dominated by nearly equal-mass binaries, with median mass ratios of $\tilde{q}\sim0.8$ and $\sim0.75$, respectively. For the post-processed population $z<4$, LISA becomes sensitive to a wider range of BBH masses at lower redshifts, including a significant number of more asymmetric systems. As a result, the median mass ratio decreases to $\tilde{q}\sim0.25$.

In the SE model, a similar trend is observed. ET and LGWA remain preferentially sensitive to nearly equal-mass binaries, whereas LISA detects a larger fraction of asymmetric systems owing to its sensitivity to more massive BBHs. However, While the detectable $z>4$ population is characterized by median mass ratios of $\tilde{q}\sim0.4$ for ET and $\sim0.3$ for LGWA, the $z<4$ population reaches median values of $\tilde{q}\sim0.9$ and $\sim0.7$, respectively. This difference is a consequence of the simplified post-processing adopted for systems merging below $z=4$, in which the BH masses are kept fixed after the end of the \texttt{CAT} simulation and no further accretion is assumed during the remaining pairing phase. As a result, the differential BH growth predicted by the SE accretion model during this phase is not accounted for, leading to larger mass ratios for the post-processed delayed-merger population.

The inclusion of the post-processed lower-z candidates naturally shifts the detectable population toward lower redshifts, resulting in generally higher median S/N than for binaries merging within the original \texttt{CAT} redshift range.  
This reflects the reduced luminosity distances of the population together with the different binary mass distributions probed by the three detectors.
The total detection rates across $0<z<24$ for ET, LGWA, and LISA, respectively, are $85.65 \pm 12.00$ yr$^{-1}$, $31.44 \pm 4.86$ yr$^{-1}$, and $39.51 \pm 0.91$ yr$^{-1}$ in the EL model, and $75.95 \pm 2.59$ yr$^{-1}$, $59.55 \pm 8.73$ yr$^{-1}$, and $100.38 \pm 6.16$ yr$^{-1}$ in the SE scenario.

%%%%%%%%%%%%%%%%%%%%%%%%%%%%%%%%%%%%%%%%%%%%%%%%%%%%%%%%%%%%%%
Figure~\ref{fig:waterfall_lower_z} shows the redshift distribution of detectable merging BBHs as a function of the total mass in the source-frame, for the entire detectable population at redshift $z>0$. As in Fig.~\ref{fig:waterfall}, hexagonal bins are color-coded according to the number density of BBH mergers ($\rm Mpc^{-3}$), while the colored areas are the ET (red), LGWA (blue), and LISA (green) waterfall plots for equal-mass binaries ($q\simeq1$). To facilitate comparison with detector sensitivities, only nearly equal-mass systems ($q \geq 0.7$) of \texttt{CAT} are shown.

In the EL model, the $z<4$ population is concentrated in two distinct mass ranges, around $\sim10^2$ -- $10^4\,\rm M_\odot$ and $\sim10^5$ -- $10^7\,\rm M_\odot$. The gap between $10^4$ and $10^5\,\rm M_\odot$ is a direct consequence of the discrete BH seeding prescription discussed in Sect.~\ref{sec:discussion}. The low-mass population falls primarily within the sensitivity ranges of ET and LGWA, whereas only the high-mass component is accessible to LISA.

Compared to the EL case, the SE model exhibits a broader and more continuous distribution in the low-mass end, from $\sim10^2$ to $\sim10^7\,\rm M_\odot$. The more efficient BH growth characteristic of the SE accretion model populates the intermediate- and high-mass regime more effectively, filling the gap present in the EL model. As a result, the $z<4$ population spans the sensitivity ranges of all three detectors: ET and LGWA remain sensitive to the low and intermediate-mass systems, whereas LISA probes a substantially larger population of massive mergers than in the EL model.

\vspace{5mm}

\end{appendix} 

\end{document}